\documentclass[a4paper,11pt]{article}
\pdfoutput=1 

\usepackage{jheppub} 

\usepackage[T1]{fontenc} 
\usepackage{amsmath,mathtools}
\usepackage{mathrsfs}
\usepackage[shortlabels]{enumitem}
\usepackage{amsthm}
\usepackage{float}

\newcommand{\R}{\mathbb R}

\newcommand{\Vol}{\mathrm{Vol}}
\newcommand{\aether}{\ae ther\;}

\DeclareUnicodeCharacter{2212}{-}

\title{\boldmath Mixed Gauge-Global Symmetries, Elliptic Modes, and Black Hole Thermodynamics in Ho\v{r}ava-Lifshitz Gravity}

\author[a]{L. Martin}
\author[a]{and D. Mattingly}

\affiliation[a]{University of New Hampshire, USA}

\emailAdd{luke.martin@unh.edu}
\emailAdd{david.mattingly@unh.edu}

\abstract{In Ho\v{r}ava-Lifshitz gravity, a putative consistent theory of quantum gravity for which there is evidence for both black hole thermodynamics and a holographic construction, spacetime is endowed with a preferred dynamical spacelike foliation. The theory has a leaf reparameterization symmetry that is neither global nor local gauge, hyperbolic and elliptic equations of motion, a lack of splittability, and universal horizon black hole solutions. The reparameterization symmetry is ``mixed'': it is a local symmetry in one coordinate yet global on each leaf.  More broadly it is an example of both unfree and projectable gauge symmetries. 
The mixed symmetry and associated charge has not yet been accounted for in calculations of universal horizon thermodynamics in Ho\v{r}ava-Lifshitz gravity.  This has led to problems, in particular the failure of the first law in a class of asymptotically AdS solutions where the normal to the leaves of the foliation is not aligned with the time translation Killing vector at infinity. 
We show how the dynamics of the charge corresponding to this symmetry coupled with the other features above resolves this issue.  We then briefly comment how this mixed symmetry, the corresponding charge, and the elliptic equations of motion also conspire to evade recent holographic arguments for only local gauge fields in consistent theories of quantum gravity due to the lack of splittability of the elliptic equation and associated mode.}

\makeatletter
\gdef\@fpheader{}
\makeatother

\begin{document} 
\maketitle
\flushbottom

\section{Introduction}\label{sec:introduction}
One of the most important characteristics of any physical theory is the symmetries imposed on the action.  Requiring local gauge invariance is an integral part of the standard model of particle physics, local Lorentz symmetry dictates much of the form of special relativity, and the active diffeomorphism invariance of general relativity is a key component of holography and the AdS/CFT correspondence. As a consequence, symmetries also play a key role in the physics of causal horizons. The first law of black hole mechanics can be viewed as an application of Noether's theorem, the spacetime metric is affected by the gauge field charges on the black hole, etc.  In this paper we discuss the consequences for black hole thermodynamics of a ``mixed'' symmetry, which is neither a global nor local gauge symmetry, that exists in Ho\v{r}ava-Lifshitz gravity, a putative complete theory of quantum gravity.  Ho\v{r}ava-Lifshitz gravity also possesses black holes~\cite{UH} and strong indications of a good thermodynamics of the corresponding universal horizons~\cite{Mechanics, toward_thermo, Stefano, StefanoVisser, raytracing} for flat and Lifshitz asymptotics. The mixed symmetry has not so far been explicitly examined in universal horizon thermodynamics, however. Additionally, the thermodynamics for asymptotically AdS has proven problematic, in that the first law is not consistent for all asymptotically AdS spacetimes. 

The existence of a mixed symmetry and a failure of black hole thermodynamics in AdS may immediately raise concerns about the overall consistency of Ho\v{r}ava-Lifshitz gravity. Standard lore in quantum gravity, especially in the context of AdS/CFT, states that there are no global symmetries in a consistent quantum theory of gravity, but rather only local gauge symmetries \cite{Harlow_Main,Harlow_PRL}. Hence one might suspect that the failure of black hole thermodynamics in AdS is related to the existence of this mixed symmetry, which in turn may indicate that Ho\v{r}ava-Lifshitz is not actually a consistent quantum gravity theory but instead belongs to the swampland.  We indeed find that the mixed symmetry and the prior failures of black hole thermodynamics in AdS are intimately related.  However, instead of being an indication of inconsistency, the mixed symmetry, associated charge, and other seemingly unrelated features of Ho\v{r}ava-Lifshitz gravity all conspire to fix the current approaches to black hole thermodynamics when properly accounted for.  While the focus of this paper is on black hole thermodynamics, a side conclusion is therefore that in some quantum gravity theories mixed symmetries may actually play a key role instead of signaling problems. 

We now briefly explain the three key features of Ho\v{r}ava-Lifshitz gravity that make this possible to provide a conceptual picture for the reader and return to each in detail in later sections. 


\textbf{Feature: Foliations and reparameterization invariance.} Ho\v{r}ava-Lifshitz gravity~\cite{Horava_Lifshitz, Horava_Membranes} is a theory of gravity where spacetime is endowed with a dynamical spacelike foliation, the leaves of which are labelled by a scalar field $T$, the khronon.  The existence of this foliation allows one to write a theory of gravity that is Lifshitz in the ultraviolet and ghost-free~\cite{Horava_Lifshitz, Horava_Membranes}.  The ultraviolet Lifshitz symmetry of the action then permits the theory to be renormalizable~\cite{Blas}, which yields a diffeomorphism invariant and seemingly complete and well-behaved theory of quantum gravity.  The cost of the renormalizability of this theory is thus exact Lorentz invariance; there are tight experimental constraints~\cite{multimessenger, Kostelecky}, but these can be theoretically evaded via the right hierarchies and renormalization group flow~\cite{Pospelov, Status}.  The structural difference between Ho\v{r}ava-Lifshitz gravity and general relativity is the existence of the foliation.  Crucially, only the foliation is added to the theory - the actual labels of the foliation, the values of $T$, are irrelevant.  This introduces a symmetry in the theory: the theory must be invariant under monotonic reparameterizations of $T$, i.e. $T \rightarrow f(T)$ is a symmetry as long as $f(T)$ is monotonic.  This heavily restricts the form of the action.  The reparameterization symmetry is not a global symmetry - the corresponding global symmetry would be a shift symmetry $T \rightarrow T+ T_0$ where $T_0$ is a constant.  However, it is also not fully local since $f(T)$ depends only on $T$ and not on the position on any one leaf of the foliation; it is, in the parlance of Ho\v{r}ava-Lifshitz gravity, projectable.  We will consider in this paper the reparameterization symmetry as a partially gauged global symmetry, local in $T$ but global along each leaf, and the ramifications.  We denote such a symmetry as a \textit{mixed gauge-global symmetry}, or \text{mixed} symmetry for short.  There is a corresponding charge, but this charge does not source any fully local gauge field.  


\textbf{Feature: Universal Horizon thermodynamics} 
Ho\v{r}ava-Lifshitz possesses causally trapped surfaces known as universal horizons with seemingly good thermodynamics~\cite{Mechanics, toward_thermo, Stefano, StefanoVisser, raytracing, Pacilio:2017emh, Pacilio:2017swi,DelPorro:2023lbv}. The ultraviolet Lifshitz symmetry dictates that ultraviolet field excitations travel infinitely fast.  This does not mean that causality is lost; signals still must propagate to future of each leaf.  However, relativistic causality is meaningless in Ho\v{r}ava-Lifshitz gravity.  As a result, the causal horizons are not the usual black hole horizons of general relativity - for static spacetimes the event horizon is not a Killing horizon.  Instead, the dynamics of the leaves of the foliation is such that they can ``bend over'' in static spacetimes with a central mass.  While at infinity the timelike Killing vector $\chi$ in such a spacetime can be orthogonal to the leaves, if the central mass is concentrated enough the bending of the leaves will always be sufficient so that at some finite radius inside the Killing horizon the Killing vector becomes tangential to the leaves.  At this radius, since every signal propagates to the future in the foliation, a trapped surface and hence causal boundary is formed: the universal horizon.  This universal horizon has a good thermodynamics and shows promise for a full holographic description with boundary Lifshitz field theories~\cite{Horava_Lifshitz, Horava_Membranes, Iranian, CFTno, wedges}.  Key for the compatibility of mixed symmetries, their charges, and black hole thermodynamics is that all leaves outside the foliation asymptote to the universal horizon for static spacetimes and extend to spatial infinity on the other side.

\textbf{Feature: The presence of an elliptic mode}  The requirement of a foliation coupled with the fully dynamical nature of all fields leads to an additional equation of motion for the lapse $N$, which is itself a function of $T$.  Importantly, this is an elliptic equation of motion.  Unlike mathematically similar constraint equations, such as the Gauss constraint, this elliptic equation is not automatically preserved by the Hamiltonian; it is not a constraint.  Rather, it must be imposed as an actual equation of motion on every leaf~\cite{Henneaux}.  The existence of an elliptic mode in a theory with black holes seems immediately problematic, as elliptic modes couple boundary data on the ends of any leaf, and if the leaf penetrated any causal boundary then one would be able to probe any central singularity by looking at the boundary data on the other side.  This would violate assumed properties such as cosmic censorship.  However, \textit{because} all the leaves outside a universal horizon asymptote to it, the boundary data for the leaf couples points at spatial infinity not to central singularity but instead to the universal horizon data itself.  In this way the elliptic mode provides a new, unique mechanism by which boundary data at a causal horizon, but not inside, can be transmitted to spatial infinity.  It is in this way that the charge associated with this new symmetry can theoretically be emitted from the universal horizon despite there being no local gauge field to generate a chemical potential for different local charged, emitted particles~\cite{Hook:2014mla}.  We note that this behavior is \textit{only} possible with a) the imposition of Lifshitz symmetry (which was originally required for renormalizablility) which makes the universal horizon the causal horizon and b) foliation dynamics, which is required for holography and also generates the elliptic mode needed to pull charge off a causal boundary without a local gauge field.  Specific to holographic constructions, the mixed symmetry is possible because the elliptic mode violates the criterion of \textit{splittability}~\cite{Harlow_Main,Harlow_PRL} used in current proofs.

The three features above combine into a scenario where there is a mixed symmetry which is global along leaves of the foliation yet gauged in (khronon) time, a corresponding conserved charge which sources no local gauge field, causal boundaries and black hole thermodynamics, and an elliptic mode mechanism that allows horizon data and the charge to be transmitted to infinity without gauge fields.  This is a novel extension of black hole thermodynamics with qualitatively new symmetry constructions.\footnote{Of note here is the recent surge of interest in generalized gauge theories~\cite{Gaiotto, Harlow_Main,Heckman:2024oot} - for an introduction see~\cite{Gen_Sym_Review}.  The shift operator maps leaves into themselves, i.e. it acts intrinsically on extended objects rather than local ones.  Whether the corresponding reparameterization symmetry can be embedded into the framework of gauged generalized global symmetries is currently not known. 
For the purposes of this paper, when we refer to global symmetries we will not be considering generalized global symmetries, but rather the usual ``0-form'' global symmetries that can act on local objects.}  

The above picture, while perhaps conceptually compelling, is a large amount of work to quantitatively check and verify all aspects.  In the rest of this paper we concentrate on a simple first calculation to establish the relevance of the mixed gauge-global charge for black hole thermodynamics in Ho\v{r}ava-Lifshitz theory.  
We show that explicitly requiring the mixed gauge-global charge to be conserved fixes the prior problem with asymptotically AdS spacetimes - non-alignment of the normal vector corresponds to an external flux that changes the mixed gauge-global charge.  Requiring a first law between equilibrium solutions for all charges therefore removes this possibility and explains why the existing derivations of thermodynamics for universal horizons failed in this case. 

The structure of the paper is as follows. In Section~\ref{sec:HL} we introduce Ho\v{r}ava-Lifshitz gravity, including the basic foliation structure, the relation to Einstein-aether theory, action, and notation.  Of the three necessary features,  the elliptic mode is introduced in~\ref{sec:elliptic}, reparameterization invariance as a mixed symmetry and the corresponding charge is discussed in Section~\ref{sec:symmetries}, while the third, universal horizons and their thermodynamics, is explained in Section~\ref{sec:UHthermo}.  We then show how the existing thermodynamics fails in the asymptotically AdS case in Section~\ref{sec:AdSfail}.  Resolving this failure using the three features is done in Section~\ref{sec:resolution}, and finally we conclude with some remarks and future directions regarding the usual narratives around global/gauge symmetries and splittability in Section~\ref{sec:conclusion}.  

\section{Basics of Ho\v{r}ava-Lifshitz Gravity}\label{sec:HL} 

\subsection{Foliations}

In Ho\v{r}ava-Lifshitz gravity space and time are treated on different footings due to the existence of the dynamical spacelike foliation. While this foliation is dynamical, rather than a gauge choice, the formalism is in exact analogy to an ADM 3+1 decomposition in general relativity.  The spacetime is split into spacelike hypersurfaces $\Sigma$, or leaves, via a \textit{foliation} diffeomorphism $\phi$,

\begin{equation}\label{foliation dif}
\phi: \mathcal{M}\to \R\times S.
\end{equation}
Here, $S$ in the codomain represents space and $\R$ is the temporal direction. 
The pullback of the time function $t:\R\times S\to \R$ by the foliation diffeomorphism gives a natural time coordinate on $\mathcal{M}$,  $T=\phi^*t$, which is called the ``khronon''.
If we then consider the vector field $\partial_t$ on $\R\times S$, its clear that pushing this vector field forward by $\phi^{-1}$, we get a timelike vector field on the full spacetime $\mathcal{M}$, namely $\partial_T=\phi^{-1}_*\partial_t$. It is then useful to specify the components of this timelike vector field with respect to the leaves of the foliation, namely, an ADM style normal and a tangential component.

The components of the timelike vector field $\partial_T$ are given by
\begin{equation}\label{partial T decomp}
\partial_T=(N,\Vec{N})
\end{equation}
where the component normal to $\Sigma$, $N$, is the usual lapse, and the component tangent to $\Sigma$, $\Vec{N}$, the shift. This allows us to write the metric in terms of the foliation in the following fashion:
\begin{equation}\label{adm metric}
ds^2=-N^2dT^2+p_{ij}(dx^i+N^idt)(dx^j+N^jdt)
\end{equation}
where $p_{ij}$ represents the induced 3-metric on a given leaf $\Sigma$. 

\subsection{Action and low energy equations of motion}
Given the existence of the foliation, one specifies the low energy dynamics by constructing the most general two derivative action invariant under foliation-preserving diffeomorphisms. The ultraviolet dynamics would then be specified by higher spatial derivative terms, the form of which won't concern us here except that they be Lifshitz.  The most general infrared action one can write down to second order under these conditions while leaving the lapse $N$ a dynamical function on $\mathcal{M}$ is~\cite{Horava_Lifshitz,Horava_Membranes}
\begin{equation}\label{HL action}
S=\frac{M_{pl}^2}{2}\int d^3x\, dt N\sqrt{p} \bigg[K^{ij}K_{ij}-\lambda K^2+\xi \prescript{(3)}{}{R}+\eta a_ia^i\bigg].
\end{equation}
Here, $K_{ij}$ the extrinsic curvature of a leaf which can be expanded as
\begin{equation}\label{extrinsic curvature}
K_{ij}=\frac{1}{2N}(\dot{p}_{ij}-\nabla_i N_j-\nabla_j N_i)
\end{equation}
$\prescript{(3)}{}{R}$ is the 3-Ricci scalar associated with the 3-metric $p_{ij}$, and $\lambda, \xi, \eta$ are naively free parameters. Finally, 
\begin{equation}\label{acceleration dlnN}
a_i=\partial_i \ln N
\end{equation}
is the acceleration of the unit normal to the leaves in the preferred foliation \cite{Blas}. The acceleration $a_a$ was not included in the initial formulation of Ho\v{r}ava-Lifshitz gravity, but was rather included later in non-projectable Ho\v{r}ava-Lifshitz gravity developed in \cite{Blas}, a covariant formulation of Ho\v{r}ava gravity where the lapse, $N=N(t,x)$, is a function of the full spacetime rather than just of time. Specifically, one can write $N$ in terms of the khronon as

\begin{equation}\label{lapse fn of T}
N^{-2}=-g^{ab}(\nabla_a T)(\,\nabla_b T)
\end{equation}

In order to ensure that the dynamics of the theory are well defined as pointed out in \cite{Henneaux}, one must add to the action terms constructed from the 3-vector $a_i$ \cite{Blas}.  Note that due to the reparameterization invariance there is no direct $T$ coupling in the action, only derivative terms.

It is often convenient to define the following tensor
\begin{equation}\label{G tensor}
G^{ijkl}=\frac{1}{2}\big(p^{ik}p^{jl}+p^{il}p^{jk}\big)-\lambda p^{ij}p^{kl}
\end{equation}
and its inverse when $\lambda\ne 1/3$:
\begin{equation}\label{G inverse}
\mathcal{G}_{ijkl}=\frac{1}{2}(p_{ik}p_{jl}+p_{il}p_{jk})-\frac{\lambda}{3\lambda-1}p_{ij}p_{kl}.
\end{equation}
With this tensor, we can rewrite \eqref{HL action} as:
\begin{equation}\label{np action G rewrite}
S=\int d^3x dt N\sqrt{\gamma} \big(G^{ijkl}K_{ij}K_{kl}+\xi\prescript{(3)}{}{R}\big)
\end{equation}

\subsection{Relationship to Einstein-Aether theory and notation}\label{subsec:aether equiv sect}

Due to the absence of direct khronon couplings, the Ho\v{r}ava-Lifshitz action can be written in terms of the vector field on $\mathcal{M}$ constructed from the unit normal to the leaves $\Sigma_T$. In the IR limit, Ho\v{r}ava-Lifshitz gravity is equivalent to twist-free Einstein-\aether theory \cite{aether, Universal}. The equivalence between these two theories was proven in \cite{aether_Horava}. 

In its original formulation, Einstein-\aether theory is a generally-covariant theory where the dynamical fields consist of the spacetime metric $g_{ab}$, and the ``\ae ther'' timelike vector field $u^a$. The action of this theory can be expressed as
\begin{equation}\label{aether action}
S=\frac{1}{16\pi G_{\text{\ae}}}\int_\mathcal{M} \Vol_\mathcal{M}\,\left[-\frac{6c_{cc}}{l^2}+R+\mathcal{L}_{\mathrm{\text{\ae}}}+\mathcal{L}_{\text{\ae}}^{(CON)}\right]
\end{equation}
where
\begin{equation}\label{aether constraint}
\mathcal{L}_{\text{\ae}}^{(CON)}=\lambda_{\text{\ae}}(u^2+1)
\end{equation}
constrains the \ae ther field $u_a$ to be timelike and to have unit norm, and $c_{cc}=0,\pm 1$ along with the length scale $l$ determines the curvature of the spacetime where $c_{cc}=0$ corresponds to asymptotically flat spacetimes, $c_{cc}=1$ to asymptotically de Sitter, and $c_{cc}=-1$ to asymptotically anti de Sitter.

The kinetic part of the \aether Lagrangian $\mathcal{L}_{\text{\ae}}$ is given by
\begin{equation}\label{aether kinetic}
\mathcal{L}_{\text{\ae}}=-Z^{ab}_{\;\;\;\;cd}(\nabla_a u^c)(\nabla_b u^d)
\end{equation}
where
\begin{equation}\label{aether Z}
Z^{ab}_{\;\;\;\;cd}=c_1 g^{ab}g_{cd}+c_2\delta^a_{\;c}\delta^b_{\;d}+c_3\delta^a_{\;d}\delta^b_{\;c}-c_4u^a u^b g_{cd}
\end{equation}
ensures all possible two-derivative terms of the \aether field. 

The \textit{twist} 3-form is defined as $\omega=u\wedge du$ and is constructed out of the \aether $u\in\Omega^1(\mathcal{M})$ which is most naturally a 1-form on $\mathcal{M}$. One can see in the case that $u=-NdT$, as is the case when the \aether is ``orthogonal'' to the leaves of the preferred foliation, the twist is clearly vanishing due to the antisymmetry of the wedge product. From this point forward, the twist 3-form will be assumed to vanish since we will only deal with a hypersurface-orthogonal \ae ther.

The relation between these two theories in the IR limit becomes clear if we note the following identities
\begin{equation}\label{aether identities 1}
\nabla_a u_b=-u_a a_b+K_{ab}\hspace{2cm}p_{ab}=u_a u_b+g_{ab}
\end{equation}
where $g_{ab}$ again denotes the metric on the full spacetime. Additionally, the components of the \aether are taken to be 
\begin{equation}\label{u components}
u_a=-N\nabla_a T.
\end{equation}
It is clear then that \eqref{aether constraint} is consistent \eqref{lapse fn of T}. Additionally, the acceleration can be also be written in terms of the \aether as
\begin{equation}\label{accel aether}
a_a=u^b\nabla_bu_a.
\end{equation}
Finally, with the couplings $c_i$, we define the following quantities:
\begin{align}\label{c defns}
c_{13}=c_1+c_3\hspace{1.5cm}c_{14}=c_1+c_4\hspace{1.5cm}c_{123}=c_2+c_{13}.
\end{align}
which allows the coupling constants in the two theories to be related via
\begin{equation}\label{constants dictionary}
\xi=\frac{1}{1-c_{13}}\hspace{1.5cm}\lambda=\frac{1+c_2}{1-c_{13}}\hspace{1.5cm}\eta=\frac{c_{14}}{1-c_{13}}.
\end{equation}

The key difference between the IR limit of the covariant formulation of Ho\v{r}ava-Lifshitz gravity and Einstein \aether theory is the treatment of the khronon, $T$, as the fundamental scalar field of the theory in the former versus the treatment of the \aether vector field $u^a$ as the fundamental field in the latter. However, as demonstrated in \cite{aether_Horava}, the black hole solution space of twist-free Einstein-\aether theory is a proper subset of that of Ho\v{r}ava-Lifshitz gravity. It was then demonstrated in \cite{Universal} that the two solution spaces are equivalent in the case of staticity and spherical symmetry.  The equations of motion, in \aether notation are given by \cite{Universal, Mechanics}:
\begin{equation}\label{eoms}
\mathcal{G}_{ab}=T_{ab}\hspace{1.5cm} \text{\AE}_a=0
\end{equation}
where
\begin{equation}\label{eom specifics}
\begin{aligned}
T_{ab}=&\lambda_{\text{\ae}}u_au_b+c_4a_aa_b-\frac{1}{2}g_{ab}Y^c_{\;\;d}\nabla_cu^d+\nabla_c X^c_{\;\;ab}\\
&+c_1\left[(\nabla_a u_c)(\nabla_b u^c)-(\nabla^c u_a)(\nabla_c u_b)\right]\\
\text{\AE}_a=&\nabla_b Y^b_{\;\;a}+\lambda_{\text{\ae}}u_a+c_4(\nabla_a u^b)a_b\\
Y^a_{\;\;b}=&Z^{ac}_{\;\;bd}\nabla_c u^d\\
X^c_{\;\;ab}=&Y^c_{\;\;(a}u_{b)}-u_{(a}Y_{b)}^{\;\;c}+u^c Y_{(ab)}
\end{aligned}
\end{equation}
As usual, $\mathcal{G}_{ab}$ denotes the Einstein tensor, and $T_{ab}$ denotes the stress tensor associated with the \ae ther.  These are the equations of motion that must be solved for the black hole solutions in Section~\ref{sec:UHthermo}.

\section{Feature : Elliptic mode in Ho\v{r}ava-Lifshitz Gravity}\label{sec:elliptic}

One of the unique features of Ho\v{r}ava-Lifshitz gravity is the so-called ``elliptic mode'', which arises since space and time are treated on different footings in a dynamical manner. The equations of motion contain an elliptic equation that allows one to solve for the lapse function, $N$, on the entirety of a given slice $\Sigma_T$ when given the appropriate boundary data. This means that elliptic mode not only needs to be taken into account when on-shell reducing various quantities (which will be done in subsequent sections), but also has important ramifications regarding black hole thermodynamics.

The elliptic mode is most clearly understood as being born out of the constraint algebra. While it is not a proper constraint under the Dirac formalism, it is a condition that must be satisfied by the lapse to enforce closure of the algebra. We detail this construction below.  In the GR limit, the constraint algebra closes automatically and the need for an elliptic condition on the lapse function vanishes.

For this section, we will chose the notation of the canonical, Hamiltonian formulation of Ho\v{r}ava-Lifshitz gravity rather than that of the covariant Einstein-\aether theory. The mixed symmetry of Ho\v{r}ava Lifshitz gravity is most easily understood in the canonical formalism since it arises due to the different treatments of space and time in the action. Additionally, the canonical formulation is the most natural for studying this constraint algebra. How the mixed symmetry and elliptic charge can be understood more broadly in the context of covariant phase space methods of \cite{Wald_Phase, Carlip, Harlow_Phase, Tony} is an open question beyond the scope of this paper.

The review in this section of the constraint algebra and elliptic mode will initially closely follow \cite{Consistency} while additionally including the acceleration term to ensure a well behaved theory at low energies \cite{Blas}. After demonstrating how the elliptic mode emerges in asymptotically flat spaces, we will then generalize to more general asymptotics by adding the bare cosmological constant to the Hamiltonian (a more detailed review of asymptotics in Ho\v{r}ava-Lifshitz gravity will be covered in Section \ref{Asymptotics Section} below).

As in general relativity, we take the 3-metric of a leaf $p_{ij}$ to be the generalized position coordinate. In the case of Ho\v{r}ava gravity, we take this 3-metric to be that of a leaf in the preferred foliation. It can be shown that the momentum conjugate to $p_{ij}$ is given as \cite{Consistency}:
\begin{equation}\label{conj mom HLnp}
\pi^{ij}=\sqrt{p}\,G^{ijkl}K_{kl}
\end{equation}
Making use of \eqref{extrinsic curvature} and performing a Legendre transformation, one can find that the Hamiltonian density for the theory is given by
\begin{equation}\label{H dens tot}
\mathcal{H}_{tot}=N\mathcal{H}+N_i\mathcal{H}^i
\end{equation}
where
\begin{equation}\label{hamiltonian components}
\begin{aligned}
\mathcal{H}=&\frac{\pi^{ij}\pi^{kl}}{\sqrt{p}}\mathcal{G}_{ijkl}-\sqrt{p}\;(\xi\prescript{(3)}{}{R}+\eta a_ia^i)\\
\mathcal{H}^i=&-2N_i\nabla_j\pi^{ij}=\pi^{ij}_{\;\;;\,j}.
\end{aligned}
\end{equation}
The total Hamiltonian is unsurprisingly obtained by integrating \eqref{H dens tot} over a given leaf of the foliation, namely:
\begin{equation}
H=\int_{\Sigma}d^3x\, \mathcal{H}_{tot}
\end{equation}
As in general relativity, the Einstein field equations imply the vanishing of $\mathcal{H}$ and $\mathcal{H}^i$, and thus also $H$. The constraint $\mathcal{H}=0$ is referred to as the \textit{Hamiltonian constraint}, also in keeping with the convention of general relativity.

Recall that to enforce preservation in time of the primary constraint $\mathcal{H}$, it is sufficient to compute its Poisson bracket with itself. Preservation in time of a constraint $\phi$ amounts to enforcing that $\{\phi,H\}=0$; however, since $\mathcal{H}^i$ generates spatial diffeomorphisms, Poisson brackets between $\mathcal{H}^i$ and $H$ correspond to infinitesimal coordinate transformations and lead to no new constraints \cite{Consistency,DeWitt}. Thus the only remaining term in the bracket $\{\mathcal{H},H\}$ is $\{\mathcal{H},\mathcal{H}\}$.

Proceeding along these lines, we find that the Poisson bracket $\{\langle A \mathcal{H}\rangle,\langle B \mathcal{H}\rangle\}$, where $A,B$ are in general distributions and the angled brackets denote integration over $\Sigma$, is given by
\begin{equation}\label{H poisson general}
\{\langle A \mathcal{H}\rangle,\langle B \mathcal{H}\rangle\}=2\int d^3x\, d^3y\, A \; \pi^{ij}\bigg[\nabla_i\nabla_j B -\bigg(\frac{\lambda- 1}{3\lambda-1}\bigg)g_{ij}\nabla^2 B\bigg]-(A\leftrightarrow B)
\end{equation}
Now we can take $A=\delta(x-y)$ and $B=N$ the lapse and integrate by parts to obtain 
\begin{equation}\label{H,NH}
\{\mathcal{H},\langle N\mathcal{H}\rangle\}=2\bigg(\frac{\lambda-1}{3\lambda-1}\bigg)\big(N\nabla^2\pi+2\partial_iN\nabla^i\pi\big)+2\mathcal{H}^i\partial_i N+N\nabla_i\mathcal{H}^i
\end{equation}
Recalling that $\mathcal{H}^i=0$, we find that enforcing $\{\mathcal{H},\langle N\mathcal{H}\rangle\}=0$ is equivalent to enforcing that
\begin{equation}
N\nabla^2\pi+2\partial_iN\nabla^i\pi=0
\end{equation}
Since $N\ne0$, it is clear that this constraint is equivalent to requiring 
\begin{equation}\label{momentum constraint pre int}
\nabla_i(N^2\nabla^i \pi)=0
\end{equation}
Multiplying this quantity by $\pi/\sqrt{p}$ and integrating by parts over $\Sigma$, we obtain 
\begin{equation}
\int_\Sigma \Vol_\Sigma \,N^2(\nabla_i\pi)^2=0
\end{equation}
Again, since $N^2>0$ on the entirety of $\Sigma$, this is equivalent to  
\begin{equation}\label{pi difeq}
\nabla_i\pi=0
\end{equation}

The only regular solution of \eqref{pi difeq} that satisfies zero momentum at the asymptotic boundary is the second-class constraint
\begin{equation}\label{pi=0}
\pi=0
\end{equation}
One can then obtain the desired elliptic mode by enforcing the preservation in time of the constraint in \eqref{pi=0} which once again amounts to computing the Poisson bracket of the constraint with $\mathcal{H}$, as per the above discussion. One then finds
\begin{equation}
\{\pi,\langle N\mathcal{H}\rangle \}=-2\sqrt{p}\xi(\vec{\nabla}^2-\prescript{(3)}{}{R})N+2\eta N\sqrt{p}\, a_ia^i+\frac{3}{2}N\mathcal{H}
\end{equation}
Noting again that $\mathcal{H}=0$, we find that enforcing that $\pi=0$ be preserved in time amounts to requiring that the lapse $N$ satisfy the following elliptic equation:
\begin{equation}\label{elliptic mode}
 \left(\vec{\nabla}^2-\prescript{(3)}{}{R}-c_{14}\,a_ia^i\right)N=0.
\end{equation}
Note that~\eqref{elliptic mode} is exactly as found in \cite{Consistency} with the addition of the acceleration term. Here, we have made use of \eqref{constants dictionary} in simplifying \eqref{elliptic mode}.

The next natural step, then, is to generalize \eqref{elliptic mode} to more general asymptotics as will be necessary to our study of asymptotically AdS universal horizons. One can do this rather simply by adding a cosmological constant term to the Hamiltonian
\begin{equation}\label{H redef}
\begin{aligned}
\Tilde{\mathcal{H}}=&\mathcal{H}+\mathcal{H}_\Lambda\\
\mathcal{H}_\Lambda=&\sqrt{p}\xi\frac{6c_{cc}}{l^2}.
\end{aligned}
\end{equation}
Since $\mathcal{H}_\Lambda$ contains no derivatives, it is clear that
\begin{equation}\label{H lambda with self}
\{\langle A\Tilde{\mathcal{H}}\rangle,\langle B\Tilde{\mathcal{H}}\rangle\}=\{\langle A\mathcal{H}\rangle,\langle B \mathcal{H}\rangle\}.
\end{equation}

One might note that the constraint $\pi=0$ must be modified since it was obtained through the assumed boundary condition for $\pi_{ij}=0$ at infinity which only holds in asymptotically Minkowski spacetimes. However, since $K$ must be constant at infinity \cite{Universal}, and since the $\pi=0$ constraint is reached by solving $\nabla_i\pi=0$ on a given hypersurface and applying the asymptotically flat boundary condition $\pi_{ij}=0$, if we enforce the boundary condition $\pi=\pi'$ some constant at infinity, then we simply enforce in time the constraint $\tilde{\pi}=\pi-\pi'=0$. Since $\pi'$ is a constant, it is clear that the bracket will be unaffected as it vanishes upon differentiation. The only bracket we must therefore compute is 
\begin{equation}\label{pi H lambda}
\{\tilde{\pi},\langle N\mathcal{H}_\Lambda\rangle\}=\{\pi,\langle N\mathcal{H}_\Lambda\rangle\}=-\xi\int_\Sigma \Vol_\Sigma\frac{9c_{cc}}{l^2}.
\end{equation}
Since the bracket is linear, we have
\begin{equation}\label{pi H tilde}
\{\pi,\langle N\Tilde{\mathcal{H}}\rangle\}=-\int_\Sigma\Vol_\Sigma 2\sqrt{p}\left[\xi\left(\vec{\nabla}^2-\prescript{(3)}{}{R}+\frac{9c_{cc}}{l^2}\right)-\eta a^2\right]N-\frac{3}{2}N\Tilde{H}.
\end{equation}
Thus, by enforcing the equations of motion, which now amounts to saying that $\Tilde{\mathcal{H}}=0$, we can see that the elliptic equation for $N$ with general asymptotics is given by
\begin{equation}\label{elliptic mode general}
\left[\xi\left(\vec{\nabla}^2-\prescript{(3)}{}{R}+\frac{9c_{cc}}{l^2}\right)-\eta a^2\right]N=0.
\end{equation}
In \aether notation, this reads

\begin{equation}
\left[\left(\vec{\nabla}^2-\prescript{(3)}{}{R}+\frac{9c_{cc}}{l^2}\right)-c_{14} a^2\right]N=0.
\end{equation}

\section{Feature : Reparameterization invariance and conserved charge}\label{sec:symmetries}

Ho\v{r}ava-Lifshitz gravity is invariant under reparameterization symmetry, $T \rightarrow f(T)$ as long as $f(T)$ is monotonic.  As we saw, this constrained the form of the action.  The reparameterization invariance can be rewritten as $T \rightarrow T + \tilde{f}(T)$ without loss of generality as long as monotonicity is preserved.  In coordinates adapted to the foliation, $T$ simply defines the coordinate time $t$ and so this transformation becomes a spacetime time translation $t \rightarrow t+\tilde{f}(t)$ where the translation parameter is now local in time.  

Theories with such a time (and space) dependent translations can be constructed as localizations of theories with global translation symmetries.  For example, typical non-relativistic free theories possess global Galilean invariance, where the theory is invariant under $x^a \rightarrow x^a + \xi^a$ where $\xi^a$ is a constant.  There is a well defined localization procedure for such non-relativistic theories \cite{diffR} where $\xi^a$ becomes coordinate dependent, leading to a gauge theory with a three dimensional diffeomorphism invariance and time reparameterization invariance.  We take the same approach here, simply in the foliation labels $T$, which are equivalent to a time coordinate in the foliation frame.  However, unlike a full localization in both space and time of a globally Galilean invariant theory only the time localization is needed to match the reparameterization invariance.  The reparameterization invariance is then a mixed, or partially localized/gauged invariance that corresponds to the global symmetry $T \rightarrow T + T_0$, i.e. a global foliation translation/shift symmetry.  Clearly this is a global transformation, preserves the foliation structure, and is a symmetry of the action.

As an aside on nomenclature, functions that are functions of only $T$ are called ``projectable'' in Ho\v{r}ava-Lifshitz gravity~\cite{Horava_Membranes, Horava_Lifshitz}.  Indeed, originally projectability was a key aspect of the lapse in the initial construction, although this was later shown to be too restrictive~\cite{Blas}.  Hence one could also think of mixed symmetries as ``projectable'' symmetries.  However, at the same time the reparameterization transformation is an example of an ``unfree'' gauge transformation~\cite{unfree}.  Unfree gauge transformations are those where the gauge parameter is not a local function of $t,x$ but instead satisfies some differential equation.  In the case of Ho\v{r}ava-Lifshitz gravity, the differential equation would simply be $\vec{\nabla} \tilde{f}(x,T)=0$, where $\vec{\nabla}$ is the spatial gradient along a leaf.  This, up to a meaningless global shift, restricts $\tilde{f}(x,T)$ to simply $\tilde{f}(T)$, as we have been discussing.  Finally, as mentioned previously, whether this symmetry can be embedded into the framework of generalized global symmetries is unknown.  We choose the term ``mixed'' in this paper to emphasize the dual global/gauge nature of the transformation without unnecessarily pigeonholing it into any of the existing conceptual frameworks it is related to.

Given the existence of the above mixed and corresponding global symmetry in Ho\v{r}ava theory, Noether's theorem dictates a conserved quantity. This is easily determined in the usual fashion by requiring that the action \eqref{HL action} be stationary under variations in $T$ and collecting boundary terms, which we now do.

Using the identities found in Section \ref{subsec:aether equiv sect}, one can easily determine the following variations
\begin{align}
\delta_Tu_a=&-Np_{ab}(\nabla^b\delta T)\label{important variation 1}\\
\delta_T N=&-u_a N^2(\nabla^a\delta T)\label{important variation 2}\\
\delta_T a_a=&-N\,K_{ab}(\nabla^b\delta T)-u_c\nabla^c(N\,p_{ab})(\nabla^b\delta T)\label{important variation 3}\\
&-u_cNp_{ab}(\nabla^c\nabla^b\delta T)\nonumber
\end{align}
where $\delta_T$ denotes variation with respect to $T$. In the usual fashion, one obtains
\begin{equation}\label{action variation general}
\delta_T S=\int_\mathcal{M} \left(d\star J+E_T\cdot \delta T\right)
\end{equation}
where $E_T$ denotes the equation of motion corresponding to $T$, and $J$ the current corresponding to variations in $T$. After a few lines of work and use of the identities in \eqref{important variation 1}-\eqref{important variation 3}, one finds that the components of the current $J$ are given by
\begin{equation}\label{current components}
J_b=\delta T\big(f_b-\nabla^a(\Tilde{f}_{ab}+\Tilde{f}_{ba})\big)
\end{equation}
where
\begin{equation}\label{current components explicit}
\begin{aligned}
f_b=&-u_b N\mathcal{L}-c_{13}\big[-(-u_a a_b+K_{ab})(\nabla^aN)-N(u_bK_{ac}K^{ac})\\
&-Na^cK_{bc}-u_aa_b\nabla^a N-Nu_ba^2\big]\\
&-2c_2 K\big[-NK u_b-Na_b-\nabla_bN\big]+2c_{14}\big[-Na^cK_{bc}-u_aa_b\nabla^aN-Nu_ba^2\big]\\
\nabla^af_{ab}=&2c_{13}\big[K_{ab}\nabla^aN+N\nabla^aK_{ab}\big]+2c_2K\big[\vec{\nabla}_bN+Nu_bK+Na_b\big]\\
&-2c_{14}\big[u_aa_b\nabla^aN+u_aN\nabla^aa_b+Na_bK\big]\\
\nabla^af_{ba}=&2c_{13}\big[K_{ab}\nabla^aN+N\nabla^aK_{ab}\big]+2c_2K\big[\vec{\nabla}_bN+Nu_bK+Na_b\big]\\
&-2c_{14}\big[u_ba_a\nabla^aN+u_bN\nabla^aa_a+Na_a\nabla^au_b\big]
\end{aligned}
\end{equation}

Using this current, one can thus obtain the corresponding charge by integrating the current $\star J$ over a slice $\Sigma$:
\begin{equation}\label{QT defn}
Q_T=\int_\Sigma\star J_T.
\end{equation}
%
%
%
Using \eqref{current components explicit}, one finds that the charge reads
\begin{equation}\label{charge pre on shell}
Q_T=\int_\Sigma d^3x \sqrt{p}\,N\left[\mathcal{L}+2c_{13} K_{ab}K^{ab}+2c_2K^2-2c_{14}a^2-2c_{13}a^2-c_{14}\nabla_aa^a\right]
\end{equation}
where $\mathcal{L}$ denotes the Lagrangian density. One notes that in the GR limit where $c_i\to 0$ for all $i$, $\mathcal{L}$ reduces to the Einstein-Hilbert Lagrangian, and the remaining terms of \eqref{charge pre on shell} vanish. After an on-shell reduction using $H=0$ (we have dropped the tilde from here on out for simplicity, but it is assumed that $H$ here is the Hamiltonian for general asymptotics), and \textit{additionally} the elliptic mode equation from section \ref{sec:elliptic}, one finds that the charge, which we will call the \emph{elliptic charge}, is given by
\begin{equation}\label{QT on shell}
Q_T=\int_\Sigma d^3x\, \sqrt{p}\left[2\left(\frac{1}{1-c_{13}}\right)\vec{\nabla}_i(Na^i)+2N\frac{c_{123}}{1-c_{13}}K^2-2c_{13}Na^2-N\frac{15c_{cc}}{(1-c_{13})\,l^2}\right].
\end{equation}
Note that the first term of \eqref{QT on shell}, which does not vanish in the general relativistic limit, can be understood as a consequence of how the foliation approaches asymptotic infinity. Essentially, it is similar to a non-orthogonal corner term~\cite{Hawking_Corner, Odak:2021axr} where there is also a mismatch between the asymptotic Killing vector and the normal to the foliation. To see the foliation geometry more clearly, one can rewrite this first term as
\begin{equation}\label{Q 1 rewritten}
I_1=\int_\Sigma d^3x\sqrt{p}\,\vec{\nabla}_a(Na^a)=\int_{\partial\Sigma}d^2x\sqrt{\hat{g}}\,(Ns_aa^a)=\int_{\partial\Sigma}d^2x\sqrt{\hat{g}}N s_a u_b\nabla^b u^a
\end{equation}
Integrating by parts, one finds
\begin{equation}\label{Q1 int by parts}
I_1=-\int_{\partial\Sigma}d^2x \sqrt{\hat{g}}\,Nu^au^b\nabla_as_b=-\int_{\partial\Sigma}d^2x \sqrt{\hat{g}}\,N(p^{ab}-g^{ab})\nabla_as_b.
\end{equation}

From the second equality in \eqref{Q1 int by parts}, it is clear that this term measures the difference in extrinsic curvature of the surface defined by $s_a$ at infinity when measured with respect to the Ho\v{r}ava-Lifshitz foliation versus the spacetime as a whole. Namely, if the three and four divergences of $s_a$ are equal, $I_1$ is clearly vanishing.  In general relativity, this term does not play as important of a role since one could always chose a foliation where this term would vanish. However, in Ho\v{r}ava Lifshitz gravity, one must work with the preferred foliation, which in general will introduce non-orthogonality of leaves $\Sigma_T$ with the timelike boundary $\partial\mathcal{M}$, and in turn introducing non-orthogonal corner-type terms such as $I_1$. Further understanding of how such corner-type terms could affect this charge and the thermodynamics of universal horizons in general using the methods of covariant phase space explored in \cite{Wald_Phase, Carlip, Harlow_Phase, Tony, Odak:2021axr} will be left for future work.

Additionally, by examining \eqref{QT on shell}, one can see that changing the required boundary data to solve the elliptic mode, namely $\nabla_iN$ and $N$, will make a corresponding change to the charge $Q_T$, hence the the name ``elliptic charge''. Since Ho\v{r}ava-Lifshitz gravity admits black hole solutions, the existence of such a charge raises questions about the form of the first law governing said black holes.  First, in a truly static solution that corresponds to a time independent equilibrium thermodynamical state, this charge should be conserved.   Second, if this charge gravitates, as in the case of the electromagnetic charge of a Reissner-Nordstrom black hole for instance, the charge will appear in the black hole's first law for the appropriate ensemble. In the following section, we cover the formulation of black holes in Ho\v{r}ava-Lifshitz gravity, and the difficulties of formulating first laws for universal horizons in asymptotically AdS. We will then see that use of the elliptic charge \eqref{QT on shell} resolves some of the known issues.

\section{Feature : Universal horizon thermodynamics}\label{sec:UHthermo}
At first glance, Ho\v{r}ava-Lifshitz gravity would seem to admit no causal horizons.  The ultraviolet mode behavior is Lifshitz, which implies that the group velocity of wave modes diverges at high energy.  This precludes any finite limiting speed for all modes, which would seem to forbid the existence of a causal horizon. Contrary to this expectation, Ho\v{r}ava-Lifshitz gravity has been shown to exhibit causal horizons known as \emph{universal horizons}. In essence, in a static situation with a timelike Killing vector $\chi^a$ and associated Eddington-Finkelstein Killing time coordinate $v$, the bulk dynamics of the foliation leaves themselves allow for solutions where the leaves all asymptote to a leaf not connected to spacelike infinity, denoted as the universal horizon, for infinite negative $v$.   Since all modes, even those with infinite group velocity, must propagate to the future in khronon time, the universal horizon acts as a trapped surface and a causal horizon, as we detail below. 

 A rigorous discussion of these black holes can be found in \cite{Causality}. In addition, maximally symmetric universal horizon solutions have been studied in \cite{Universal}, which will largely be reviewed in this section. An important result of \cite{Universal} is that the static, spherically symmetric black hole solution spaces of Einstein-\aether theory and Ho\v{r}ava-Lifshitz are equivalent. Therefore, we will largely use coupling constants and parameters from Einstein-\aether theory for simplicity.  

\subsection{Universal Horizons in general}\label{UH Section}

 For analyzing static and spherically symmetric universal horizon spacetimes we will follow the presentation in~\cite{Jishnu}, and refer the reader there for additional details.  We work in Eddington-Finklestein (EF) coordinates, in which the metric takes the form

\begin{equation}\label{EF metric}
ds^2=-e(r)dv^2+2f(r)dvdr+r^2d\Omega_2^2.
\end{equation}
Note that in these coordinates, the timelike Killing vector is $\chi^
\alpha=\partial_v$ with $e(r)=-\chi^2$. Along with $\chi$, it will be useful to define the radial 1-form 
$\rho_a=f(r)dr$ which is orthogonal to $\chi^a$, namely $\chi\cdot\rho=0$. The components of $\chi^a$ and $\rho_a$ and those of the dual 1-form and vector field (respectively) are given by \cite{Universal, Jishnu}
\begin{equation}\label{chi and rho}
\chi^a=\partial_v\,,\hspace{0.5cm}\rho^a=\partial_v+\frac{e(r)}{f(r)}\partial_r\,,\hspace{0.5cm}\chi_a=-e(r)dv+f(r)dr\,,\hspace{0.5cm}\rho_a=f(r)dr.
\end{equation}

It will be algebraically useful to expand quantities along unit vectors aligned with the leaves, rather than expand along $\chi$ and $\rho$. We define the unit vector parallel to the aether acceleration, $s^a \propto a^a$, which is everywhere orthogonal to $u$ and by definition lies within a given slice $\Sigma_T$.  Given such a unit vector, we have 
\begin{equation}\label{s recall}
a_a=(a\cdot s)s_a\hspace{1.5cm}s^2=1\hspace{1.5cm} (u\cdot s)=0.
\end{equation}
By spherical symmetry, one can expand various quantities in the $u,s$ basis. A useful way of expanding the extrinsic curvature, $K_{ab}$, of a slice $\Sigma_T$ will thus be
\begin{equation}\label{spherical Kab}
\nabla_au_b=-(a\cdot s)u_as_b+K_{ab}\hspace{1.5cm} K_{ab}=K_{ss}s_as_b+\frac{\hat{K}}{2}\hat{g}_{ab}
\end{equation}
where $\hat{g}_{ab}$ denotes the metric induced on the 2-sphere, and $\hat{K}$ denotes the trace of $K_{ab}$ over the two sphere, namely $\hat{g}^{ab}K_{ab}=\hat{K}$. The full trace of the extrinsic curvature is given by
\begin{equation}\label{spherical trace K}
K=K_{ss}+\hat{K}.
\end{equation}
The timelike Killing vector for the static spacetime, $\chi$, can be expanded in the $u,s$ basis:
\begin{equation}\label{chi in u s}
\chi^a=-(u\cdot \chi)u^a+(s\cdot \chi)s^a.
\end{equation}
This allows one to define the ``surface gravity''~\cite{StefanoVisser} through
\begin{equation}\label{kappa defining}
\nabla_a\chi_b=-\kappa \epsilon_{ab}^{II}
\end{equation}
where $\epsilon_{ab}^{II}=u_as_b-s_au_b$, and the surface gravity, $\kappa$, is thus
\begin{equation}\label{surf grav defn}
\kappa=-(a\cdot s)(u\cdot \chi)+K_{ss}(s\cdot \chi).
\end{equation}

In our chosen coordinate system, using e.g. \eqref{chi and rho} along with the selected orientation of $u$ and $s$ given by $(u\cdot\chi)=-(s\cdot\rho)$ and $(s\cdot\chi)=-(u\cdot\rho)$, one finds that the \aether can be expressed as 
\begin{equation}\label{aether in EF}
u=(u\cdot\chi)dv+\frac{f(r)dr}{(s\cdot\chi)-(u\cdot\chi)}
\end{equation}
where we can additionally note that $e(r)=(u\cdot\chi)^2-(s\cdot\chi)^2$, which ensures that the unit norm condition of the \aether is satisfied \cite{Universal, Jishnu}.
By recalling that $u$ is assumed to be hypersurface orthogonal, we should be able to easily extract the behavior of the surfaces of constant $T$ from \eqref{aether in EF} using $u=-NdT$. 
This requires us to note that $N\propto (u\cdot\chi)$, and is equal up to some function of the khronon. More precisely, $N=f(T)(u\cdot\chi)$. If we gauge fix $T$ via $\chi^\alpha\nabla_\alpha T=1$, we can see that \cite{Universal}
\begin{equation}\label{Lapse as u dot chi}
N=-(u\cdot\chi).
\end{equation}

Using \eqref{Lapse as u dot chi} and hypersurface orthogonality, one can immediately see from \eqref{aether in EF} that
\begin{equation}\label{dT}
dT=dv+\frac{f(r)dr}{(u\cdot\chi)\left[(s\cdot\chi)-(u\cdot\chi)\right]}.
\end{equation}
From \eqref{dT}, it is clear that our leaves $\Sigma_T$ in EF coordinates will be given by integral curves of the vector field defined by \cite{Universal, Jishnu}
\begin{equation}\label{const T difeq}
\frac{dv}{dT}=-\frac{f(r)}{(u\cdot\chi)\left[(s\cdot\chi)-(u\cdot\chi)\right]}
\end{equation}
which are plotted in EF coordinates in Figure \ref{leaves} below. We see then that if there is a leaf where $u \cdot \chi=0$, the EF time coordinate $v$ is along the leaf itself, i.e. the leaf never reaches spatial infinity. $(u\cdot\chi)=0$ defines a universal horizon if we require that matter fields follow future directed causal curves with respect to the foliation~\cite{Universal}. Given the understanding of the geometry and how $u \cdot \chi=0$ defines a universal horizon, we now turn to the global backgrounds that solutions with universal horizons are asymptotic to, and then the solutions themselves.
\begin{figure}[h]
\centering
\includegraphics[width=0.7\linewidth]{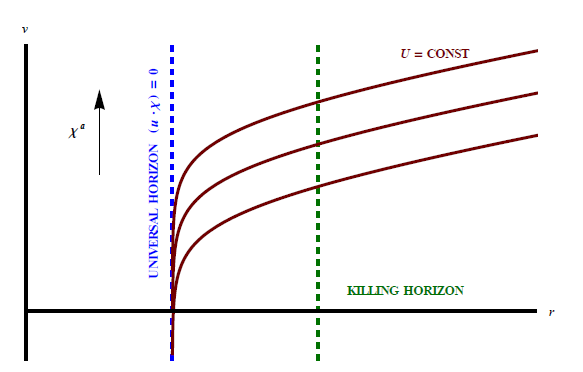}
\caption{Leaves $\Sigma_T$ outside of the universal horizon (depicted by brown curves above) determined by \eqref{const T difeq} asymptote to the universal horizon at $r_{UH}$ such that $(u\cdot\chi)=0$.  Reproduced from \cite{Universal}.}
\label{leaves}
\end{figure}

\subsection{Globally Maximally Symmetric Solutions}

Since we will want black hole solutions which are smoothly connected to the ``vacuum'', i.e. the appropriate globally maximally symmetric solutions, we wish for solutions to have the following asymptotic behavior \cite{Universal}:
\begin{equation}\label{desired asymptotics}
e(r)\sim -\frac{\Lambda r^2}{3} + 1 + \mathcal{O}(r^{-1})\hspace{1.5cm} f(r)\sim 1 +\mathcal{O}(r^{-1}).
\end{equation}
Here, $\Lambda$ is the effective cosmological constant, generated due to the bare cosmological constant and the aether background.  In general it is not equal to the bare cosmological constant found in the action, namely $\frac{6c_{cc}}{l^2}$; however, one can be expressed in terms of the other using parameters describing the alignment of the \aether with the Killing vector $\chi$ at infinity. These parameters will be introduced in this section, and the relationship of the bare and effective cosmological constant will be introduced in the following section.

Ensuring that black hole solutions are ultimately smoothly connected to these globally maximally symmetric solutions allows for an appropriate background for the thermodynamics.   Restricting our study to solutions which are are smoothly connected to the maximally symmetric solutions given in \eqref{desired asymptotics} is equivalent to only considering solutions that can be obtained as excitations of these ground states, which will allow for an appropriate background subtraction.

It turns out that while only approximate solutions can be found in the case where $f(r)\ne 1$, one can find \textit{all} analytic black hole solutions with $f(r)=1$ \cite{Universal}, which is the restriction we will make for this and subsequent sections.
We begin by noting that in the $u,s$ basis (which we will henceforth assume due to the assumed spherical symmetry), the off-diagonal components of the Ricci tensor are given by \cite{Universal}
\begin{equation}\label{Rus}
R_{us}=\frac{2(s\cdot \chi)(u\cdot \chi)f'(r)}{r\,f(r)^3}=0
\end{equation}
where the second equality follows from our assumption that $f(r)=1$. Thus, since $R_{us}=0$, it follows from Einstein's equation that \cite{Universal}
\begin{equation}\label{equations for uh}
\begin{aligned}
&\frac{c_{14}(s\cdot\chi)}{r^2}\left[r^2(u\cdot \chi)'\right]'=0\\
&c_{123}(u\cdot\chi)\left[r^{-2}[r^2(s\cdot \chi)]'\right]'=0
\end{aligned}
\end{equation}
The second equation in \eqref{equations for uh} follows from another consequence of Einstein's equation in this case, which is that \cite{Universal}
\begin{equation}
c_{123}\nabla_s K=0.
\end{equation}
This should come as no surprise to the reader since we saw in Section \ref{sec:elliptic} that in the case of general asymptotics, $\pi\propto K$ must be a constant to ensure closure of the constraint algebra of the theory.

From \eqref{equations for uh}, one can then obtain the desired globally maximally symmetric solutions. However, in order to do this, one must first choose constraints on the coupling constants since, for example, global AdS solutions do not exist for generic couplings, namely $c_{123}\ne 0$ and $c_{14}\ne 0$, nor for $c_{123}=0$ and $c_{14}\ne 0$. Rather, the only choice of couplings that gives a globally AdS solution is for $c_{14}=0$ and $c_{123}\ne 0$. Note that one cannot simultaneously set $c_{123}=c_{14}=0$, as this would eliminate the \aether equation of motion in the case of Einstein \aether theory, and the equation of motion for the lapse, $N$, in Ho\v{r}ava Lifshitz gravity \cite{Universal}. The foliation would hence be pure gauge and not physically dynamical.
In the following subsections, we will present these solutions for the various restrictions on the coefficients.

\subsubsection{Globally Anti de-Sitter}

As noted above, the only choice of couplings that admits a globally anti de-Sitter solution is $c_{14}=0$ and $c_{123}\ne 0$. As one can see from \eqref{equations for uh}, and from the fact that $e(r)=(u\cdot \chi)^2-(s\cdot\chi)^2$, specifying a solution amounts to obtaining expressions for $(u\cdot\chi)$ and $(s\cdot\chi)$. Thus, the only globally AdS solution allowed by \eqref{equations for uh} is \cite{Universal}

\begin{equation}\label{global ads solns}
(u\cdot\chi)=-\sqrt{1+\frac{r^2}{l_u^2}}\hspace{1.5cm}(s\cdot\chi)=\frac{r}{l_s}
\end{equation}
with $l_u$ and $l_s$ related by
\begin{equation}\label{l conditions global ads}
l_u^2=\frac{l^2l_s^2}{c_l l^2-c_{cc}l_s^2}\hspace{2cm}l_u<l_s
\end{equation}
where $c_l=\frac{1}{2}(2+3c_2+c_{13})$.  $l_s$ is the only free parameter and it controls the alignment of $u$ with $\chi$, for $l_s \rightarrow \infty$ $s$ is always orthogonal to $\chi$ and $u$ is always parallel. In the global solutions these parameters control alignment everywhere. However, as we relax our view to solutions that only asymptotically approach these forms, these parameters will only specify the alignment at spatial infinity. This will be covered in greater detail in Section \ref{Asymptotics Section} below.

Global AdS will end up being the most important to understand since asymptotically AdS universal horizons will be the only black holes of interest to us in later sections. Our ultimate goal in this paper is to understand more completely the thermodynamics of universal horizons in the context of the novel mixed symmetry possessed by Ho\v{r}ava-Lifshitz gravity. In previous work, first laws have been obtained for solutions of universal horizons in asymptotically flat black holes \cite{Universal, Mechanics}, as will be laid out in Section \ref{mechanics section} below. However, when one relaxes the conditions on asymptotics to allow asymptotically AdS spacetimes, one finds that there is difficulty in obtaining a coherent first law for corresponding universal horizons. It is for this reason that asymptotically AdS universal horizons will be of primary interest in this paper. However, for the sake of completeness, we present the remaining global solutions.

\subsubsection{Globally Minkowski}

Globally Minkowski space is a solution for all couplings. For $c_{14}\ne 0$, the solutions are given by
\begin{equation}\label{global mink c14ne0}
(u\cdot\chi)=-1\hspace{2cm}(s\cdot\chi)=0
\end{equation}
with no free parameters. If one sets $c_{14}=0$, one then has
\begin{equation}\label{global mink c14=0}
(u\cdot\chi)=-\sqrt{1+\frac{r^2}{l_s^2}}\hspace{1.5cm}(s\cdot\chi)=\frac{r}{l_s}
\end{equation}
where the alignment parameters are constrained by
\begin{equation}
c_{cc}l_s^2=(c_l-1)l^2.
\end{equation}
$l_s$ is a free parameter if and only if $c_{cc}=0$ and $c_l=1$. Otherwise, once again, there are no free parameters.

\subsubsection{Globally de-Sitter}

Finally, we present the globally de-Sitter solutions, which again exist for all allowed combinations of coupling constants. For $c_{14}\ne 0$, they are
\begin{equation}\label{global ds c14ne0}
(u\cdot\chi)=-1\hspace{2cm}(s\cdot\chi)=\frac{r}{l_{\text{eff}}}
\end{equation}
where 
\begin{equation}\label{leff}
l_{\text{eff}}=
\begin{cases}
l\sqrt{c_l}&c_{123}\ne 0\\
l\sqrt{1-c_{13}}&c_{123}=0
\end{cases}
\end{equation}
and there are no free parameters. Alternatively if $c_{14}=0$, we have that
\begin{equation}\label{global ds solns}
(u\cdot\chi)=-\sqrt{1+\frac{r^2}{l_u^2}}\hspace{1.5cm}(s\cdot\chi)=\frac{r}{l_s}
\end{equation}
with
\begin{equation}\label{l conditions global ds}
l_u^2=\frac{l^2l_s^2}{c_l l^2-c_{cc}l_s^2}\hspace{2cm}l_u<l_s
\end{equation}
which is exactly the case for global AdS, with the exception of the fact that $l_u>l_s$.

\subsection{Asymptotic structure of spacetimes containing universal horizons}\label{Asymptotics Section}

Universal horizon spacetimes admit multiple asymptotic behaviors with different symmetry groups.  Universal horizons were first studied in asymptotically flat, spherically symmetric spacetime~\cite{Barausse:2011pu}.  This analysis was then extended to asymptotically AdS and dS spaces, still with spherically symmetric horizons, in~\cite{Universal} of which the AdS solutions will be of the most interest to us.  Beyond AdS, planar universal horizons have been found and studied with Lifshitz asymptotics~\cite{Basu:2016vyz}, as globally Lifshitz spacetime is a solution of the Ho\v{r}ava-Lifshitz field equations and Lifshitz asymptotics are relevant for dual Lifshitz field theories~\cite{Griffin:2012qx, wedges}.  

Our specific example by which we show the effect of mixed symmetries will be the spherically symmetric AdS case, as this exhibits the necessary physics and analytic solutions have been found.  In this scenario the asymptotic behavior of the maximally symmetric universal horizon solutions studied in \cite{Universal} are controlled by four parameters.  The first, and obvious parameter is the usual cosmological constant term which appears in the action as $-\frac{6c_{cc}}{l^2}$, where $c_{cc}=0,\pm 1$. As stated above, $c_{cc}=0$ corresponds to Minkowski space, whereas $\pm 1$ correspond to de-Sitter and anti de-Sitter geometries, and the parameter $l$ is the length scale associated with the curvature. 

In general relativity, where the only geometric degree of freedom is the metric, specifying the approach of the metric to AdS in a Fefferman-Graham construction suffices. 
In Ho\v{r}ava-Lifshitz gravity however additional information must be specified: in addition to the usual asymptotic metric geometry, one must specify the asymptotic geometry of the leaves in the dynamically preferred foliation.  The asymptotic geometry of the foliations is controlled by the quantities $(u\cdot\chi)$ and $(s\cdot\chi)$ at infinity, or alignment with the Killing vector $\chi^a$ at infinity. The length scales associated with the ``large r'' behavior of these quantities are $l_u$ and $l_s$ respectively \cite{Universal}, only one of which is independent.

In the case where $c_{14}=0$ and asymptotically AdS, the trace of the extrinsic curvature is given by \cite{Universal}
\begin{equation}\label{c14=0 K}
K=-\frac{3}{l_s}.
\end{equation}
It is clear that as $l_s\to\infty$, which corresponds to the so-called ``aligned case'', the extrinsic curvature becomes traceless, as one might expect.

 It is also useful to define an effective cosmological constant, $\Lambda$, which is not independent but can be expressed in terms of the alignment parameters $l_u$, $l_s$ along with the length scale associated with the bare cosmological constant, $l$. Specifically, \cite{Universal}
\begin{equation}\label{l relations}
\begin{aligned}
\frac{c_{cc}}{l^2}=&\frac{c_l}{l_s^2}-\frac{1}{l_u^2}\hspace{1.5cm}\frac{c_l-1}{l_u^2}=\frac{c_{cc}}{l^2}-\frac{c_l\Lambda}{3}\\
\frac{\Lambda}{3}=&\frac{1}{l_s^2}-\frac{1}{l_u^2}\hspace{1.5cm}\frac{c_l-1}{l_s^2}=\frac{c_{cc}}{l^2}-\frac{\Lambda}{3}.
\end{aligned}
\end{equation}

\subsection[Maximally Symmetric Black Hole Solution with c14=0]{Maximally Symmetric Black Hole Solution with \(c_{14}=0\)}

Recall that spherically symmetric, static AdS solutions only exist for $c_{14}=0$. As found in \cite{Universal}, the solution for universal horizons with $c_{14}=0$ can be expressed in general as

\begin{align}\label{c14=0 general solns}
&(u\cdot\chi)=-\frac{r}{l_u}\left(1-\frac{r_{uh}}{r}\right)\sqrt{1+\frac{2r_{uh}}{r}+\frac{3r_{uh}^2+l_u^2}{r^2}+\frac{2r_{uh}(3r_{uh}^2+l_u^2)}{3r^3}+\frac{r_{uh}^2(3r_{uh}^1+l_u^2)}{3r^4}}\nonumber\\
&(s\cdot\chi)=\frac{r}{l_s}+\frac{r_{uh}^2}{r^2\sqrt{3(1-c_{13})}}\sqrt{1+\frac{3r_{uh}^2}{l_u^2}}\nonumber\\
&e(r)=-\frac{\Lambda r^2}{3}+1-\frac{r_0}{r}-\frac{c_{13}r_{uh}^4}{3(1-c_{13})r^4}\left(1+\frac{3r_{uh}^2}{l_u^2}\right)\\
&f(r)=1\nonumber\\
&r_0=\frac{4r_{uh}}{3}+\frac{2r_{uh}^3}{l_u^2}+\frac{2r_{uh}^3}{l_s\sqrt{3(1-c_{13})}}\sqrt{1+\frac{3r_{uh}^2}{l_u^2}}\nonumber
\end{align}

Here, $r_{uh}$ denotes the radial location of the universal horizon. Note that the solution presented in \eqref{c14=0 general solns} holds in general for $c_{14}=0$ and any choice of asymptotics and that one can select the asymptotics of the black hole by choosing the appropriate value for $\Lambda$ in the expression for $e(r)$ and imposing the proper constraints on $l_u,\,l_s,\,\&\,l$. A new parameter introduced to these solutions is $r_0$, which similar to the familiar Schwarzschild black hole is understood as the mass (up to a factor of 2) of the universal horizon. This connection will be made clear in later sections.  Additionally, one can note that in the limit that $r_{uh}\to 0$, one recovers the desired \eqref{global ads solns}. In other words, as desired, \eqref{c14=0 general solns} is smoothly connected to the appropriate ground state given in \eqref{global ads solns}. 

\subsection{Mechanics of Universal Horizons}\label{mechanics section}

The mechanics of universal horizons were studied early on in \cite{Mechanics} for special cases. Namely, the universal horizons for which first laws were formulated consisted of spherically symmetric and asymptotically flat geometries with \aether alignment at infinity with first $c_{123}=0$, and then $c_{14}=0$.  This was later expanded on for different asymptotics and rotating solutions in~\cite{Pacilio:2017emh}. This section will closely follow \cite{Mechanics} and will serve as a introduction to prior ideas about universal horizon mechanics, with the new interpretation of the asymptotics and role of the elliptic charge discussed in the later sections of this paper.

Recall that the black hole solution spaces of Einstein \aether theory and Ho\v{r}ava-Lifshitz gravity are equivalent. Therefore, without any loss of generality, we will follow the notational convention of \cite{Universal, Mechanics} and use that of Einstein \aether theory. In this notation, we can recall that the equations of motion, namely the Einstein equations, and the equation of motion of the \aether, are given by
\begin{equation}\label{eoms recall}
\mathcal{G}_{ab}=T_{ab}\hspace{1.5cm} \text{\AE}_a=0
\end{equation}
where
\begin{equation}\label{eom specifics recall}
\begin{aligned}
T_{ab}=&\lambda_{\text{\ae}}u_au_b+c_4a_aa_b-\frac{1}{2}g_{ab}Y^c_{\;\;d}\nabla_cu^d+\nabla_c X^c_{\;\;ab}\\
&+c_1\left[(\nabla_a u_c)(\nabla_b u^c)-(\nabla^c u_a)(\nabla_c u_b)\right]\\
\text{\AE}_a=&\nabla_b Y^b_{\;\;a}+\lambda_{\text{\ae}}u_a+c_4(\nabla_a u^b)a_b\\
Y^a_{\;\;b}=&Z^{ac}_{\;\;bd}\nabla_c u^d\\
X^c_{\;\;ab}=&Y^c_{\;\;(a}u_{b)}-u_{(a}Y_{b)}^{\;\;c}+u^c Y_{(ab)}
\end{aligned}
\end{equation}
where $T_{ab}$ is understood to be the stress tensor of the \aether and $\lambda_{\text{\ae}}$ is once again the Lagrange multiplier enforcing the unit norm constraint of the \ae ther. In \cite{Mechanics}, various constraint equations are obtained by projecting these equations of motions in various directions and taking various linear combinations. However, in order to obtain the Smarr formula, in a fashion similar to \cite{Hawking_4_Laws}, the \aether equation of motion was projected along the Killing vector $\chi^a$ in order to construct a closed two form $\mathcal{F}_{ab}$:
\begin{equation}\label{F two form}
\nabla_b\mathcal{F}^{ab}=0\hspace{1.5cm}\mathcal{F}_{ab}=q(u_as_b-s_au_b)
\end{equation}
where
\begin{equation}\label{smarr q}
q=-\left(1-\frac{c_{14}}{2}\right)(a\cdot s)(u\cdot \chi)+(1-c_{13})K_0(s\cdot\chi)+\frac{c_{123}}{2}K(s\cdot\chi).
\end{equation}

Due to the resemblance between \eqref{F two form} and the source free Maxwell equations, it was concluded that by spherical symmetry and staticity that $q\propto r^{-2}$. Thus, using the asymptotic expansions of the solutions considered in \cite{Mechanics}, namely
\begin{equation}\label{mechanics asymptotic solutions}
\begin{aligned}
(a\cdot s)=&\frac{r_0}{2r^2}+\mathcal{O}(r^{-3})\\
(u\cdot \chi)=&-1+\frac{r_0}{2r}+\frac{r_0^2}{8r^2}+\mathcal{O}(r^{-3})\\
(s\cdot \chi)=&\frac{r_{\text{\ae}}^2}{r^2}+\mathcal{O}(r^{-3})\\
\end{aligned}
\end{equation}
(here $r_{\text{\ae}}$ is the $\mathcal{O}(r^{-2})$ coefficient of the 0-component of the \ae ther, and $r_0$ is as defined above) and components of the extrinsic curvature, $K=K_0+\hat{K}$:
\begin{equation}\label{K comps expansion}
\begin{aligned}
K=&\mathcal{O}(r^{-5})\\
K_0=&\frac{2r_{\text{\ae}}}{r^3}+\mathcal{O}(r^{-5})\\
\hat{K}=&-\frac{2r_{\text{\ae}}}{r^3}+\mathcal{O}(r^{-5})
\end{aligned}
\end{equation}
it was concluded that $q$ could be written as
\begin{equation}\label{q with asymptotics}
q=\left(1-\frac{c_{14}}{2}\right)\frac{r_0}{2r^2}.
\end{equation}

In keeping with \cite{Hawking_4_Laws}, the mass of a universal horizon can be measured as the acceleration felt by an asymptotic observer, namely that given by the ADM mass:
\begin{equation}
M_{ADM}=-\frac{1}{4\pi G_{\text{\ae}}}\int_{\infty}d\Sigma_{ab}\nabla^a\chi^b
\end{equation}
where $d\Sigma_{ab}=-(u_as_b-s_au_b)dA$. If one defines the surface gravity, $\kappa$, through
\begin{equation}\label{kappa defn}
\nabla_a\chi_b=-\kappa(u_as_b-s_au_b),
\end{equation}
recall that the surface gravity can be written as
\begin{equation}\label{kappa solns}
\kappa=-(a\cdot s)(u\cdot\chi)+K_0(s\cdot\chi)
\end{equation}
thereby allowing one to write the ADM mass for the solutions specified above as
\begin{equation}\label{mechanics adm mass}
M_{ADM}=\frac{1}{4\pi G_{\text{\ae}}}\int_{\infty}dA (a\cdot s)=\frac{r_0}{2G_{\text{\ae}}}.
\end{equation}

Invoking Gauss' law, it was concluded that the flux of $\mathcal{F}_{ab}$ at the boundary at infinity must be equal to that at the inner boundary given by the universal horizon which lead to the Smarr formula for the mass of universal horizons found in \cite{Mechanics} given by
\begin{equation}\label{mechanics mass}
M_{\text{\ae}}=\frac{q_{uh}A_{uh}}{4\pi G_{\text{\ae}}}
\end{equation}
where
\begin{equation}\label{M ADM ad ae relation}
M_{\text{\ae}}=\left(1-\frac{c_{14}}{2}\right)M_{ADM},
\end{equation}
$q_{uh}$ is $q$ evaluated at the universal horizon, and $A_{uh}$ is the area of the universal horizon, namely $A_{uh}=4\pi r_{uh}^2$. 

From the Smarr formula in \eqref{mechanics mass}, it is possible to acquire a first law of universal horizons by varying \eqref{mechanics mass} along the solution subspace. Due to the fact that $M_{\text{\ae}}$ depends on one dimensionful parameter, $r_0$, one can note that $r_{uh}=\mu r_0$ where $\mu$ is some constant depending on the couplings $c_i$.  This along with the fact that $A_{uh}=4\pi r_{uh}^2\propto r_0^2$ and $q_{uh}\propto r_0^{-1}$ allows one to conclude that $\delta q_{uh}A_{uh}=-\frac{1}{2}q_{uh}\delta A_{uh}$. This leads to the follow first law for asymptotically flat universal horizons with \aether alignment at infinity:
\begin{equation}\label{mechanics first law}
\delta M_{\text{\ae}}=\frac{q_{uh}\delta A_{uh}}{4\pi G_{\text{\ae}}}.
\end{equation}
One may note that $\kappa\propto q\propto r_0^{-1}$, which allows \eqref{mechanics first law} to make contact with the usual thermodynamics of black holes familiar from general relativity.

\subsection{Radiation from Universal Horizons}

One of the tensions in any theory of gravitation that has causal horizons is that by allowing matter with non-zero entropy to fall past a causal horizon, one can decrease the entropy of the exterior universe.  If the horizon later evaporates, this can then possibly violate the generalized second law of thermodynamics. This tension is resolved for causal horizons in both general relativity and Ho\v{r}ava-Lifshitz gravity by associating a thermodynamics to the causal horizon itself~\cite{Bekenstein, Hawking_4_Laws, Hawking_particle} via horizon entropy
\begin{equation}\label{GR BH entropy}
S=\frac{A}{4G_N},
\end{equation}
and thermal radiation with temperature
\begin{equation}\label{GR BH temp}
T=\frac{\kappa}{2\pi}.
\end{equation}

Recall that in the relativistic case of a stationary black hole, Hawking radiation is understood through particle pair production at the horizon, where positive energy particles radiate outward from inside the horizon (via quantum tunneling), and negative energy particles similarly travel inward from outside the horizon. 
Following the methods of \cite{Wilczek}, the tunneling method and radiation process for asymptotically flat universal horizons was laid out in \cite{toward_thermo, Stefano} with further details given in \cite{Jishnu}. It was found that these universal horizons radiate with a temperature given by \cite{toward_thermo, Stefano}
\begin{equation}\label{Tuh}
T_{uh}=\frac{(a\cdot s)_{uh}|\chi|_{uh}}{2\pi}
\end{equation}
since $\kappa$ in Ho\v{r}ava-Lifshitz gravity is proportional to $a \cdot s$ on the universal horizon. A full analysis was more recently performed in~\cite{DelPorro:2023lbv} that firmly and comprehensively established universal horizon thermodynamics.  Hence the only key difference between general relativity and Ho\v{r}ava-Lifshitz gravity is that in the latter the relevant quantities are not evaluated at the Killing horizon, but rather at the causal horizon, which does not coincide with the Killing horizon for static spacetimes as the causal structure is not dictated by the metric alone.  

Thus, universal horizons are thought to admit a good thermodynamic interpretation in the asymptotically flat case - they radiate as a blackbody with a corresponding temperature and first law~\cite{Mechanics,Stefano, DelPorro:2023lbv}. 

\section{Failure of black hole thermodynamics in asymptotically AdS}\label{sec:AdSfail}

The current understanding of universal horizon thermodynamics as laid out in the previous sections only holds for asymptotically flat spacetimes. While it was noted in \cite{raytracing, stefanoradiation} that Hawking radiation and ray tracing from causal horizons can be thought of as a local phenomenon and should not be greatly affected by the asymptotic geometry of the spacetime, radiation has not been rigorously understood in this context.

Furthermore, and more critically, there has been previous difficulty in obtaining a first law for universal horizons with more general asymptotics. Specifically, geometries which are asymptotically AdS have proven particularly difficult geometries for which to obtain a first law of universal horizons~\cite{Universal, Pacilio:2017swi}.
To illuminate the problem we will first describe the attempt at a first law in \cite{Universal} where the same prescription as \cite{Mechanics} was used. Namely, the Smarr formula as given in \eqref{smarr q} was modified to include a cosmological constant term to account for divergences in solutions with more general asymptotics:
\begin{equation}\label{ads smarr q}
q=q_{cc}(r)-\left(1-\frac{c_{14}}{2}\right)(a\cdot s)(u\cdot \chi)+(1-c_{13})K_0(s\cdot\chi)+\frac{c_{123}}{2}K(s\cdot\chi)
\end{equation}
where here,
\begin{equation}\label{qcc}
q_{cc}(r)=\frac{c_{cc}r}{l^2}.
\end{equation}

As we will see in Section \ref{sec:resolution} below, $q_{cc}(r)$ was chosen correctly, and its origins can be more properly understood with a thorough Komar mass derivation. However, while $q_{cc}(r)$ was chosen correctly, we will see that the remaining part of $q$ does not actually hold in general as it was assumed in \cite{Universal}. A careful computation of the mass using the Komar prescription shows that the methods in \cite{Universal} did not properly subtract the background above which the gravitating mass of the black hole ought to be defined. One finds, however, that with the correct universal horizon mass, the same pathology of the first law for asymptotically AdS universal horizons persists.

The previously proposed mass formula for universal horizons as obtained from \eqref{ads smarr q} in asymptotically AdS spacetimes with $c_{14}=0$ was given in \cite{Universal} by
\begin{equation}\label{Eq 57}
M=\frac{3r_{uh}}{2}+\frac{r_{uh}^3}{l_u^2}+\frac{[2(1-c_{13})-c_l]r_{uh}^2}{l_s\sqrt{3(1-c_{13})}}\sqrt{1+\frac{3r_{uh}^2}{l_u^2}}
\end{equation}
Varying this with respect to $r_{uh}$ gives the corresponding first law for maximally symmetric asymptotically AdS universal horizons:
\begin{equation}\label{original first law}
\delta M=\frac{q_{uh}\delta A}{8\pi}
\end{equation}
where 
\begin{equation}\label{quh}
q_{uh}=\frac{2}{3r_{uh}}+\frac{3r_{uh}}{l_u^2}+\frac{[2-(5c_{13}+3c_2)]}{l_s\sqrt{3(1-c_{13})}}\left[1+\frac{9r_{uh}^2}{l_u^2}\right]\left[1+\frac{3r_{uh}^2}{l_u^2}\right]^{-1}
\end{equation}

The first term can be interpreted as the usual $T\delta S$ term given by $\frac{\kappa\delta A}{8\pi G}$, and the second as the $V\delta P$ enthalpy term given by $\frac{\theta \delta \Lambda}{8\pi G}$ as demonstrated in~\cite{Traschen}. However, the final term has no clear thermodynamic interpretation \cite{Universal}. Naively, one may wonder whether this additional term could correspond to the additional novel elliptic charge in the theory which does not exist in relativistic AdS black holes. It turns out that while this does not correspond to a charge term such as in the case of the Reissner-Nordstrom black hole, the complications arising due to this term are ultimately explained by the elliptic charge.

Furthermore, note that with \aether alignment at infinity, or specifically $l_s\to \infty$ one recovers a coherent first law since the third, problematic term vanishes in this limit. This was the case considered in \cite{Mechanics} outlined in Section \ref{mechanics section}, which is the reason a reasonable first law was able to be obtained. This remains the case for the more careful construction performed below using the proper black hole mass. The reason for a well behaved first law for asymptotically AdS universal horizons will become clear in Section \ref{conserved current section} below, where the elliptic current fluxes will be investigated in the context of the proper universal horizon mass.

\section{Fixing UH-AdS thermodynamics via conserved mixed symmetry charges}\label{sec:resolution}

As outlined in the previous section, the thermodynamics of universal horizons is reasonably well-understood in the case of asymptotically flat universal horizons with \aether alignment at infinity. Historically, when attempting to generalize the thermodynamics of universal horizons Smarr formulae were modified by simply adding a term to cancel divergences introduced from different asymptotic geometries as governed by $c_{cc}$ and $l$, namely those arising from the cosmological constant. This approach is not quite adequate, however, since these attempts have assumed finite $l_s,\,l_u$ to allow for more general \aether orientation at infinity. 

In order to properly generalize the mass formula and first law, one must not only take into account the cosmological constant, but also the stress tensor corresponding to the \ae ther. Considering both of these carefully will ensure a proper background subtraction, as is required for a proper mass formula. This is achieved through the Komar mass construction, which will be reviewed in this section for the usual relativistic cases, and then applied to Ho\v{r}ava-Lifshitz gravity. We will see that the Komar mass makes contact with the understanding in \cite{Universal} that $r_0$ appearing in the solutions corresponds to the black hole mass.

\subsection{Komar Mass, AdS, and right background subtractions}\label{Komar derivation section}

The first law of black hole mechanics gives a concrete way to relate the acceleration of an observer at infinity due to the black hole, i.e. the mass, to the degrees of freedom on the inner boundary of the spacetime, i.e. the event horizon. In the case of HL gravity, our inner causal boundary is a universal horizon, which unlike in GR, does not coincide with the Killing horizon. Although this introduces some nuances, most of the usual construction of the Komar mass goes through.
This section will take a different approach than \cite{Universal}. In fact, it will arrive at a different result since we will do the background subtraction using the stress tensor in the presence of a universal horizon.

\subsubsection{Obtaining Mass Formulae from the Komar Integral Relation}\label{Komar Section}

For the purposes of introducing the Komar mass, let $\mathcal{M}$ be an asymptotically flat spacetime of dimension $D=4$ admitting a submanifold $\Sigma$ with a boundary $\partial\Sigma$ which admits a black hole in the bulk. We seek to express the gravitating mass of said black hole as experienced by some asymptotic observer. This is achieved using the Komar mass, which relates the mass as experienced at infinity to the geometry on the black hole horizon, and is obtained using the asymptotically flat Komar integral relation, which for $\chi^a$ a Killing vector is given by \cite{Komar, Traschen, Pope}
\begin{equation}\label{komar eqn flat}
\frac{1}{4\pi G}\int_{\Sigma}d\star d\chi=\frac{1}{4\pi G}\int_{\partial\Sigma}\star d\chi=0.
\end{equation}
Note that \eqref{komar eqn flat} is ultimately vanishing due to the following geometric relation:
\begin{equation}\label{Killing geometric relation}
\nabla_b\nabla^b\chi^a=-R^a_{\;c}\,\chi^c.
\end{equation}

In the vacuum case with $\mathcal{M}$ asymptotically flat, Einstein's field equations tell us that solutions will be Ricci flat, i.e. $R_{ab}=0$. Thus, \eqref{Killing geometric relation} is vanishing, and so too is \eqref{komar eqn flat}. One then obtains the mass of the black hole by noting that the boundary of $\Sigma$ consists of the asymptotic boundary appended at infinity along with with $H$, the horizon of the black hole, namely, $\partial\Sigma=H\cup \mathcal{B}_\infty$. This allows us to write \eqref{komar eqn flat} as

\begin{equation}\label{komar flat boundary}
\int_{H}\star d\chi=\int_{\mathcal{B}_\infty}\star d\chi
\end{equation}
where the integration over the infinite boundary is interpreted as the mass of the black hole \cite{Hawking_4_Laws, Traschen}, and the horizon integral gives the familiar 
\begin{equation}\label{area int horizon}
\int_{\mathcal{B}_\infty} \star d\chi=\frac{\kappa A}{4\pi G}
\end{equation}
where $\kappa$ denotes the surface gravity on the horizon $H$. Putting this together, this gives the following formula for the mass:
\begin{equation}\label{flat mass}
M=\frac{\kappa A}{4\pi G}
\end{equation}
where the right hand side of \eqref{flat mass} can be interpreted as $TS$, or the temperature times the entropy. This is then varied to obtain the first law of black hole mechanics in vacuum for $\mathcal{M}$ asymptotically flat without angular momentum \cite{Hawking_4_Laws, Traschen}:
\begin{equation}
dM=T\,dS=\frac{\kappa}{8\pi G}dA
\end{equation}

Now, let us generalize the asymptotic geometry of $\mathcal{M}$ to allow for asymptotically de-Sitter (dS) and anti de-Sitter (AdS) spacetimes determined by a cosmological constant $\Lambda$. In this case, solutions are no longer Ricci flat. Rather, we have that
\begin{equation}\label{ads Rab}
R_{ab}=\Lambda g_{ab}.
\end{equation}
\eqref{Killing geometric relation} is then no longer vanishing and instead implies that
\begin{equation}\label{ads R chi relation}
\nabla_a\nabla^a\chi^b+\Lambda\chi^b=0.
\end{equation}
Now, since $\chi$ is Killing, it must satisfy Killing's equation by definition:
\begin{equation}\label{Killing eqn}
\nabla_{(a}\chi_{b)}=0.
\end{equation}
This implies that $d\star \chi=0$, which means that $\chi$ must be an exact form by Poincare's lemma \cite{Traschen}, given by:
\begin{equation}\label{chi exact}
\chi=d\star\omega.
\end{equation}
We call $\omega$ such that \eqref{chi exact} the \emph{\textit{Killing potential}} of $\chi$. Thus, using the fact that $\chi$ is exact, along with \eqref{ads R chi relation}, we obtain the Komar integral relation for general asymptotic geometries:
\begin{equation}\label{ads komar relation}
\frac{1}{4\pi G}\int_{\Sigma}d\star d\chi+\Lambda \, d\star\omega=\frac{1}{4\pi G}\int_{\partial\Sigma}\star d\chi+\Lambda\star\omega=0.
\end{equation}

Using the definitions from the asymptotically flat case, one gets a similar mass formula but with the addition of a cosmological constant term, to be interpreted as the thermodynamic ``$PV$'' term \cite{Traschen}:
\begin{equation}
	M=\frac{\kappa}{4\pi G}A-\frac{\Theta}{4\pi G}\Lambda	
\end{equation}
This gives the following modified first law :
\begin{equation}
dM=\frac{\kappa}{8\pi G}dA+\frac{\Theta}{8\pi G}d\Lambda
\end{equation}
Now let us generalize even further to the case where there is a non-zero stress tensor in the bulk, as this is the relevant case for the universal horizon solutions.  Since we are interested in resolving some unsolved questions regarding the first law for $c_{14}=0$ AdS universal horizons, this case will be the one that is pertinent to us. Recall Einstein's equation in this case is given by:
\begin{equation}\label{Einstein Eqn}
R_{ab}-\frac{1}{2}g_{ab}R+\Lambda g_{ab}=T_{ab}.
\end{equation}
Note that this implies that \eqref{ads Rab} now becomes:
\begin{equation}\label{Ricci relation with T}
R_{ab}=\Lambda g_{ab}+T_{ab}-\frac{1}{2}g_{ab}T.
\end{equation}

This essentially modifies \eqref{ads komar relation} by the subtraction of any gravitating contribution from matter in the bulk. This will ensure that the mass formula obtained is truly that of the black hole, and not that plus separate gravitating mass in the bulk.

With this modification, the mass formula for a black hole with potentially non-flat asymptotics and matter in the bulk can be obtained from the following integral relation:
\begin{equation}\label{matter komar integral relation}
\frac{1}{4\pi G}\int_{\partial\Sigma}(\star d\chi+\Lambda\star \omega) +\frac{1}{4\pi G}\int_{\Sigma}\left(T_{ab}-\frac{1}{2}g_{ab}T\right)\chi^a\Vol_{\Sigma}^b=0
\end{equation}
where the volume element $\Vol_{\Sigma}^a$ is that on the spacelike slice $\Sigma$. Once again, we can see that the mass formula is obtained in the following fashion:
\begin{equation}\label{generic stress tensor mass formula}
M=\frac{1}{4\pi G}\int_{H}(\star d\chi+\Lambda \star \omega)-\frac{1}{4\pi G}\int_\Sigma \left(T_{ab}-\frac{1}{2}g_{ab}T\right)\chi^a\Vol_{\Sigma}^b
\end{equation}

\subsubsection[Mass Formula for c14=0 AdS Universal Horizons]{Mass Formula for $c_{14}=0$ AdS Universal Horizons}

Recall that exact solutions for universal horizons in asymptotically AdS HL gravity for $c_{14}=0$ were found in \cite{Universal}. In this section, we will obtain a mass formula for such universal horizons and the corresponding first law using the formalism laid out in Section \ref{Komar Section} above. Note again that this will differ slightly from that found in \cite{Universal} since that background subtraction was not performed in the way prescribed by the method in Section \ref{Komar Section}.

We can begin by noting the form that \eqref{generic stress tensor mass formula} in the case of HL gravity. Recall that this is given by:
\begin{equation}\label{generic stress tensor mass formula reminder}
M=\frac{1}{4\pi G}\int_{H}(\star d\chi+\Lambda \star \omega)-\frac{1}{4\pi G}\int_\Sigma \left(T_{ab}-\frac{1}{2}g_{ab}T\right)\chi^a\Vol_{\Sigma}^b
\end{equation}
where in our case, the volume form on $\Sigma$ is given by
\begin{equation}\label{vol sigma}
\Vol_{\Sigma}^a=\frac{r^2\sin\theta}{(u\cdot \chi)}u^a dr\wedge d\Omega
\end{equation}
Let us denote this bulk integration of the stress tensor terms, in this case that of the \aether field, as $I_T$. Namely:
\begin{align}\label{IT}
I_T=&\frac{1}{4\pi G}\int_{\Sigma}\left(T_{ab}^{\text{\ae}}-\frac{1}{2}g_{ab}T^{\text{\ae}}\right)\chi^a\Vol_{\Sigma}^b\nonumber\\
=&\frac{1}{4\pi G}\int_{\Sigma}\frac{r^2\sin\theta}{(u\cdot\chi)}d^3 x\left(T_{ab}^{\text{\ae}}-\frac{1}{2}g_{ab}T^{\text{\ae}}\right)\chi^au^b
\end{align}
Noting that $g_{ab}\chi^au^b=(u\cdot\chi)$, and $T_{ab}^{\text{\ae}}\chi^au^b=-(u\cdot\chi)T_{uu}^{\text{\ae}}$ since $T_{su}^{\text{\ae}}=T_{us}^{\text{\ae}}=0$ due to the assumed spherical symmetry \cite{Ted_3_page}, and $\chi^a=-(u\cdot\chi)u^a+(s\cdot\chi)s^a$, we can rewrite $I_T$ as:
\begin{align}\label{IT Simp}
I_T=&-\frac{1}{4\pi G}\int_{\Sigma}\frac{r^2\sin\theta}{(u\cdot\chi)}d^3x\,(u\cdot\chi)\left[T_{uu}^{\text{\ae}}+\frac{1}{2}T^{\text{\ae}}\right]\nonumber\\
=&-\frac{1}{4\pi G}\int_{\Sigma} d^3x\,r^2\sin\theta \left[T_{uu}^{\text{\ae}}+\frac{1}{2}T^{\text{\ae}}\right]
\end{align}
Note that the factor of $(u\cdot\chi)$ obtained by contracting with $\chi$ and $u$ conveniently cancels the divergence in $\Vol_{\Sigma}$ found at the universal horizon.

For universal horizon solutions in asymptotically AdS with $c_{14}=0$ , the stress tensor corresponding to the \aether is diagonal in the u-s basis. The non-trivial u,s components are given by \cite{Jishnu}:
\begin{equation}\label{Tuu}
-T_{uu}^{\text{\ae}}=T_{ss}^{\text{\ae}}=\frac{3(c_{13}+3 c_2)}{2 l_s^2}+\frac{c_{13}r_{uh}^4}{r^6}\left[1+\frac{3r_{uh}^2}{l_u^2}\right]
\end{equation}
whereas the trace over the two-sphere components is
\begin{equation}\label{T hat}
\hat{T}^{\text{\ae}}=\frac{15(c_{13}+3c_2)}{l_s^2}+\frac{4c_{13}r_{uh}^4}{r^6}\left[1+\frac{3r_{uh}^2}{l_u^2}\right].
\end{equation}
The full trace of the stress tensor, 
\begin{equation}\label{total T trace}
T^{\text{\ae}}=-T_{uu}^{\text{\ae}}+T_{ss}^{\text{\ae}}+\hat{T}^{\text{\ae}},
\end{equation}
can then be obtained using \eqref{Tuu} \& \eqref{T hat}:
\begin{equation}\label{total T solutions}
T^{\text{\ae}}=\frac{6(c_{13}+3c_2)}{l_s^2}-\frac{2c_{13}r_{uh}^4}{(1-c_{13})r^6}\left(1+\frac{3r_{uh}^2}{l_u^2}\right)
\end{equation}
Examining \eqref{Tuu} \& \eqref{total T solutions}, we can see that the integrand of \eqref{IT Simp} is given by:
\begin{equation}\label{IT intgrand soln}
T_{uu}^{\text{\ae}}+\frac{1}{2}T^{\text{\ae}}=-\frac{2c_{13}r_{uh}^4}{(1-c_{13})r^6}\left[1+\frac{3r_{uh}^2}{l_u^2}\right]
\end{equation}
$I_T$ can therefore be easily integrated to give the following:
\begin{align}\label{IT integrated}
I_T=&-\frac{1}{4\pi G}\int_{\Sigma}d^3x\,r^2\sin\theta\left[T_{uu}^{\text{\ae}}+\frac{1}{2}T^{\text{\ae}}\right]\nonumber\\
=&\frac{1}{G} \int_{r_{uh}}^{\infty}dr\left[\frac{2c_{13}r_{uh}^4}{(1-c_{13})r^4}\left(1+\frac{3r_{uh}^2}{l_u^2}\right)\right]\nonumber\\
=&\frac{1}{G}\left[\frac{c_{13}}{1-c_{13}}\left(\frac{2r_{uh}}{3}+\frac{2r_{uh}^3}{l_u^2}\right)\right]
\end{align}
Our next task, then, is to compute the integral at the horizon, namely:
\begin{align}\label{IH}
I_H=&\frac{1}{4\pi G}\int_{H}\star d\chi+\Lambda\star \omega\nonumber\\
=&\frac{1}{8\pi G}\int_{H} dS_{ab}\left(\nabla^a\chi^b+\Lambda\omega^{ab}\right)
\end{align}
where $dS_{ab}=\sqrt{\hat{p}}\,\epsilon^{II}_{ab}$, $\epsilon^{II}_{ab}=(u_as_b-u_bs_a)$, and $\hat{p}_{ab}$ is the induced metric on the (D-2) - sphere at the horizon. Recall that the exact solutions for universal horizons in asymptotically AdS spacetimes with $c_{14}=0$ as found in \cite{Universal} are given by:
\begin{align}\label{c14=0 ads solns}
&(u\cdot\chi)=-\frac{r}{l_u}\left(1-\frac{r_{uh}}{r}\right)\sqrt{1+\frac{2r_{uh}}{r}+\frac{3r_{uh}^2+l_u^2}{r^2}+\frac{2r_{uh}(3r_{uh}^2+l_u^2)}{3r^3}+\frac{r_{uh}^2(3r_{uh}^1+l_u^2)}{3r^4}}\nonumber\\
&(s\cdot\chi)=\frac{r}{l_s}+\frac{r_{uh}^2}{r^2\sqrt{3(1-c_{13})}}\sqrt{1+\frac{3r_{uh}^2}{l_u^2}}\nonumber\\
&e(r)=-\frac{\Lambda r^2}{3}+1-\frac{r_0}{r}-\frac{c_{13}r_{uh}^4}{3(1-c_{13})r^4}\left(1+\frac{3r_{uh}^2}{l_u^2}\right)\\
&f(r)=1\nonumber\\
&r_0=\frac{4r_{uh}}{3}+\frac{2r_{uh}^3}{l_u^2}+\frac{2r_{uh}^3}{l_s\sqrt{3(1-c_{13})}}\sqrt{1+\frac{3r_{uh}^2}{l_u^2}}\nonumber
\end{align}
From \eqref{c14=0 ads solns}, it is straightforward to see that 
\begin{equation}\label{dchi soln}
\nabla^r\chi^v=-\nabla^v\chi^r=\frac{e'(r)}{2}.
\end{equation}
Additionally, due to spherical symmetry, we take the following Killing potential \cite{Traschen, Kastor}:
\begin{equation}\label{killing pot soln}
\omega^{rt}=-\omega^{tr}=\frac{r}{3}
\end{equation}
Using $u_a$ given in \eqref{aether in EF}, along with
\begin{equation}\label{EF s}
s_a=(s\cdot\chi)dv+\frac{f(r)\,dr}{(s\cdot\chi)+(u\cdot\chi)},
\end{equation}
one can integrate \eqref{IH} to obtain
\begin{align}\label{IH soln}
I_H=\frac{1}{G}\bigg[\frac{2r_{uh}}{3(1-c_{13})}+\frac{2r_{uh}^3}{(1-c_{13})l_u^2}-\frac{r_{uh}^3}{l_s^2}+\frac{r_{uh}^2}{l_s\sqrt{3(1-c_{13})}}\sqrt{1+\frac{3r_{uh}^2}{l_u^2}}+r_{uh}^3\left(\frac{1}{l_s^2}-\frac{1}{l_u^2}\right)\bigg]
\end{align}
where the final term comes from the integration of the Killing potential on the horizon. Compiling these results, we can obtain the desired mass formula via
\begin{equation}
M=I_H-I_T.
\end{equation}

With $I_T$ and $I_H$ given by \eqref{IT integrated} \& \eqref{IH soln}, we see that the mass formula is thus
\begin{equation}\label{ads uh mass}
M=\frac{1}{G}\left[\frac{2r_{uh}}{3}+\frac{r_{uh}^3}{l_u^2}+\frac{r_{uh}^2}{l_s\sqrt{3(1-c_{13})}}\sqrt{1+\frac{3r_{uh}^2}{l_u^2}}\right].
\end{equation}
Upon examination of \eqref{ads uh mass}, one can see that it is of the same functional form as \eqref{Eq 57} in that the third term only differs by a coefficient. It is easy to see then that the first law as obtained with the proper mass given by \eqref{ads uh mass} when varying with respect to $r_{uh}$ would be of the form
\begin{equation}\label{q komar}
q_{uh}=\frac{2}{3r_{uh}}+\frac{3r_{uh}}{l_u^2}+\frac{2}{l_s\sqrt{3(1-c_{13})}}\left[1+\frac{9r_{uh}^2}{l_u^2}\right]\left[1+\frac{3r_{uh}^2}{l_u^2}\right]^{-1}
\end{equation}
which clearly contains a third term that only differs from that in \eqref{quh} by a coefficient.  However, now the first two terms are firmly established as the temperature and enthalpy terms for asymptotically AdS, as opposed to only being assumed previously.  

Now that these terms are fully understood, the third term can be properly analyzed. It implies once again that something is awry in the thermodynamics of universal horizons in asymptotically AdS spacetimes without alignment of the \aether with the time-like Killing vector at infinity. However, the first law as obtained from the correct mass formula also becomes well-behaved and coherent when considering the aligned case where $l_s\to \infty$.  We now turn to showing how this problem is resolved.

\subsection[ls, fluxes at UH and Infinity, and conserved elliptic charge]{$l_s$, fluxes at UH and Infinity, and conserved elliptic charge}\label{conserved current section} 

In the previous subsection, we saw how the first law as obtained from the proper universal horizon mass suffers from the same problem as in previous attempts. Namely, there is a third term in the first law that lacks a clear thermodynamic interpretation, but which vanishes in the case of an aligned \aether at infinity. It turns out that the poorly behaved thermodynamics in the case of an unaligned \aether at infinity is explained  by the elliptic charge calculated in Section \ref{sec:symmetries}.

Recall that the charge $Q_T$ given in \eqref{QT on shell} was obtained from the elliptic current, $J\in\Omega^1(\mathcal{M})$ which satisfies the continuity equation by definition:
\begin{equation}\label{continuity recall}
d\star J=0.
\end{equation}
In coordinates, this reads as $\nabla_aJ^a$, which if we take the coordinates of the preferred frame tells us that
\begin{equation}\label{charge density and flux}
\nabla_u\,\rho_T=\vec{\nabla}_i\,j^i
\end{equation}
where here, $\rho_T$ denotes the charge density of the elliptic charge, $Q_T$, and we have used the usual definition of the four current:
\begin{equation}
J=\rho\, dt+ j_i\,dx^i
\end{equation}
We can see then that as usual, integrating over a spatial slice $\Sigma_T$ tells us that the \ae ther-time derivative of the elliptic charge is equal to the flux of the three-current, $j_i$, over the boundary by Stokes' theorem:
\begin{equation}\label{Du Q}
\nabla_u Q_T=\int_{\Sigma_T} d_\Sigma\star j=\int_{\partial\Sigma_T}\star j.
\end{equation}
where here $d_\Sigma$ denotes the exterior covariant derivative along a slice $\Sigma$. If $Q_T$ is conserved, \eqref{Du Q} tells us that the flux of the three-current, $j$, through the boundary is vanishing. In fact, the conservation of this charge is required if we wish to do equilibrium thermodynamics and ultimately construct a first law for the universal horizons studied above. 

Now, recall from Section \ref{sec:symmetries} that the elliptic current is given by
\begin{equation}\label{current components recall}
J_b=\delta T\big(f_b-\nabla^a(\Tilde{f}_{ab}+\Tilde{f}_{ba})\big)
\end{equation}
where
\begin{align}
f_b=&-u_b N\mathcal{L}-c_{13}\big[-(-u_a a_b+K_{ab})(\nabla^aN)-N(u_bK_{ac}K^{ac})\label{fb}\\
&-Na^cK_{bc}-u_aa_b\nabla^a N-Nu_ba^2\big]\nonumber\\
&-2c_2 K\big[-NK u_b-Na_b-\nabla_bN\big]+2c_{14}\big[-Na^cK_{bc}-u_aa_b\nabla^aN-Nu_ba^2\big]\nonumber\\
\nabla^af_{ab}=&2c_{13}\big[K_{ab}\nabla^aN+N\nabla^aK_{ab}\big]+2c_2K\big[\vec{\nabla}_bN+Nu_bK+Na_b\big]\label{delfab}\\
&-2c_{14}\big[u_aa_b\nabla^aN+u_aN\nabla^aa_b+Na_bK\big]\nonumber\\
\nabla^af_{ba}=&2c_{13}\big[K_{ab}\nabla^aN+N\nabla^aK_{ab}\big]+2c_2K\big[\vec{\nabla}_bN+Nu_bK+Na_b\big]\label{delfba}\\
&-2c_{14}\big[u_ba_a\nabla^aN+u_bN\nabla^aa_a+Na_a\nabla^au_b\big]\nonumber.
\end{align}
Since we are interested in the three current, we can simply project onto a given spatial slice $\Sigma_T$ using the projector $p_{ab}$, namely:
\begin{equation}\label{projected J}
p_{ab}\,J^a=j_b.
\end{equation}
Thus, projecting \eqref{current components recall} onto $\Sigma_T$ gives
\begin{equation}
\begin{aligned}
j^a=&-2c_{13}\left[-K^{ab}\nabla_bN-2Np^{ap}\nabla^c K_{cb}-a^bK^a_{\;\;b}\right]+2c_2 K\left[-\nabla^aN-Na^2\right]\\
&+2c_{14}\left[-Na^bK^a_{\;\;b}-a^au_b\nabla^bN+Np^{ab}u_c\nabla^ca_b+NKa^a+Np^{ab}a_c\nabla^cu_b\right].
\end{aligned}
\end{equation}
Computing the flux of $j$ through the boundary thus amounts to computing
\begin{equation}\label{boundary flux j}
\int_{\partial\Sigma_T}d^2x \sqrt{\hat{g}}\,(s\cdot j)
\end{equation}

If we then specialize to our solutions, which are spherically symmetric and static, and take $c_{14}=0$ so that we can study the asymptotically AdS universal horizon solutions, the integrand becomes:
\begin{equation}\label{s dot j}
(s\cdot j)=2\left\{c_{13}\left[-\nabla^c(NK_{ss}s_c)+N\hat{K}\frac{\hat{k}}{2}\right]+c_2 K\big[-s_a\nabla^aN-N(a\cdot s)\big]\right\}.
\end{equation}
Without loss of generality for our point, we can further and temporarily specialize to the case where only $c_2\ne 0$. In this case, if we specialize to the maximally symmetric AdS universal horizon solutions given in \eqref{c14=0 ads solns}, we then obtain
\begin{equation}\label{c2 ne 0 sj}
(s\cdot j)=-\frac{6c_{2}}{l_s}(u\cdot\chi)\,(u\cdot\chi)'
\end{equation}
where here the prime denotes differentiation with respect to the EF coordinate, $r$. Recall that the boundary of a slice in the presence of a universal horizon consists of not only the boundary appended at infinity, but also the universal horizon itself, namely $\partial\Sigma_T=H\cup\mathcal{B}_\infty$. By noting that by definition, $(u\cdot \chi)=0$ at the universal horizon, or $H$, one can see that
\begin{equation}\label{horizon j flux}
\int_{H}\star j=0.
\end{equation}

However, upon a substitution of \eqref{c14=0 ads solns}, one can clearly see that 
\begin{equation}\label{infinity j flux}
\int_{B_\infty}\star j\ne 0
\end{equation}
which tells us that 
\begin{equation}\label{boundary j flux}
\int_{\partial\Sigma_T}\star j\ne 0.
\end{equation}
In other words, $\nabla_uQ_T\ne 0$, and thus the elliptic charge is not conserved for general asymptotics when moving from leaf to leaf. However, upon noting the following identities from \cite{Universal}:
\begin{equation}\label{Eq 27 universal}
\begin{aligned}
&(a\cdot s)=-\frac{(u\cdot\chi)'}{f(r)}\hspace{1.5cm}K_{ss}=-\frac{(s\cdot\chi)'}{f(r)}\\
&\hat{K}=-\frac{2(s\cdot\chi)}{rf(r)}\hspace{2.3cm}\hat{k}=-\frac{2(u\cdot\chi)}{rf(r)}
\end{aligned}
\end{equation}
we can see that $K_{ss}\sim \frac{1}{l_s}$ and $\hat{K}\sim \frac{1}{l_s}$. Therefore, we can see that for general $c_{13},c_2\ne 0$, $(s\cdot j)\to 0$ as $l_s\to\infty$. In other words, for an aligned \aether at infinity, the flux of the elliptic current through the boundary of a spatial slice is vanishing:
\begin{equation}
\lim_{l_s\to\infty}\int_{\partial\Sigma_T}\star j=\lim_{l_s\to\infty} \big(\nabla_uQ_T\big)=0.
\end{equation}
Namely, in the case where the \aether is aligned with the timelike Killing vector at infinity, the elliptic charge is conserved. This is also the limit in which the third term of \eqref{q komar} vanishes, yielding a coherent first law of black hole mechanics for asymptotically AdS universal horizons. Note that the excess flux in the misaligned case is very similar to the non-zero symplectic flux that occurs with non-orthogonal corners~\cite{Odak:2021axr}, so this result is perhaps not surprising.

\subsection{Meaning of static equilibrium solutions in the presence of elliptic modes}

Intuitively, the vanishing flux of the elliptic current at the horizon for general \ae ther alignment can be understood as the elliptic charge acting trivially on the slices in the foliation since the leaves asymptote to the universal horizon $(u\cdot\chi)=0$. However, the non-vanishing flux at infinity tells us that the system is not actually in thermodynamical equilibrium. More precisely, the non-zero time derivative of the elliptic charge in the preferred frame means that the initial definition of staticity using only bulk degrees of freedom, as explored in \cite{Universal}, is not actually sufficient to construct a coherent first law. This was seen in the failure of the first law even when carefully using the Komar mass construction. Instead of only considering bulk degrees of freedom to define static solutions as in \cite{Universal}, one must also take into account the degrees of freedom on the boundary, i.e. the alignment parameters. This is necessitated by the existence of the elliptic equation for the lapse which allows one to solve for $N$ on the entirety of a spatial slice $\Sigma_T$ when given the appropriate boundary data. 

We saw in the previous section, however, that in the aligned case where $l_s\to \infty$, the flux of the elliptic current at the boundary $\partial\Sigma_T$ is vanishing. In this case, we see that the charge $Q$ is conserved, in which case we recover the original definition of staticity and can construct a well-behaved first law. This brings into light a perhaps better understanding of the role of the parameter $l_s$, which was previously understood only as the alignment parameter of the \aether at infinity (specifically for $l_s$, the misalignment of $s$ with $\chi$ at infinity, but this is more or less equivalent to \aether alignment through \eqref{l relations}). A perhaps more appropriate way to view $l_s$ is as the parameter which regulates the net flux of the elliptic current and therefore determines whether or not the spacetime represents an equilibrium state. Staticity is therefore not solely determined by the existence of a timelike Killing vector in the case of Lorentz-violating gravitational theories, as we can see by the example given here. In this case, we can see that to define a static spacetime in the case of Ho\v{r}ava-Lifshitz gravity, one must specify the alignment at infinity by requiring that $l_s\to\infty$. 

\section{Remarks and future directions}\label{sec:conclusion}
\subsection{No global symmetries in quantum gravity and splittability}\label{sec:noglobal}
\subsubsection{Background}
Now that we have seen how the mixed symmetry resolves outstanding issues in universal horizon thermodynamics we turn to a broader view on the existence of this symmetry in a putative consistent theory of quantum gravity in the first place. A key longstanding conjecture about what symmetries are allowed in quantum gravity is that there are no exact global symmetries in any complete and consistent theory of quantum gravity~\cite{Banks_Constraints,Banks,Polchinski_Dualities,Hsin:2020mfa,Harlow_Main,Harlow_PRL}.  This conjecture originally arose from deep considerations about black hole thermodynamics~\cite{Hawking_particle} and, as befitting the importance of symmetries, is intimately connected to other major conjectures about the structure of any viable quantum gravity theory including the weak gravity conjecture (c.f~\cite{Harlow_WGC}), various completeness conjectures (c.f.~\cite{Rudelius:2020orz}), and the distance conjecture (c.f.~\cite{Cordova:2022rer}).  Proofs that there are no global symmetries are also possible in more limited holographic constructions~\cite{Harlow_PRL,Harlow_Main}, giving a clue that the conjecture is also relevant for holographic quantum gravity in general.  As well,  calculations in Euclidean quantum gravity indicate that global symmetries are not respected quantum mechanically~\cite{Banks, Strominger,KLee_Wormhole,ABBOTT1989687,COLEMAN1990387}.

Historically, the justification for the no global symmetries conjecture has been in the form of a heuristic argument regarding black holes that radiate via Hawking radiation \cite{Hawking_particle}. More recently, the conjecture has been proven rigorously in AdS/CFT by Harlow and Ooguri in \cite{Harlow_Main, Harlow_PRL} using entanglement wedge reconstruction and splittability of boundary CFTs. In this section, we will review the heuristic justification for this conjecture and introduce the tension between the conjecture and Ho\v{r}ava-Lifshitz theory.  


Consider a theory of quantum gravity that does possess a continuous global symmetry, a corresponding Noether charge, particles charged under that symmetry so that it is physically meaningful, but no corresponding local gauge symmetry and field. If black holes exist in the theory, a particle charged under the global symmetry may fall into the black hole, thereby charging the black hole itself. Note that there is no gauge field to exert a force that might prevent such a collapse due to the increasing charge on the black hole.  For a large black hole with a large total global charge $Q$ constructed from many infalling particles, there will be a corresponding large number of possible states associated with $Q$ and the black hole will have a high entropy.  Once the black hole radiates for sufficiently long, however, the entropy of the black hole will shrink, and since there is no gauge field to preferentially radiate particles of different charges the black hole cannot radiate $Q$ away.  Therefore eventually it will be too small to account for the information necessary to have a charge $Q$, violating the laws of thermodynamics~\cite{Banks}. Additionally, if it completely evaporates without radiating the charge back out then the charge $Q$ is lost, therefore leading to non-conservation.\footnote{The analysis differs somewhat for topological charges such as skyrmions, see~\cite{Dvali:2016mur,March-Russell:2002jiy} for a discussion.}

As a corollary, one can consider what is necessary for the black hole to radiate charge away.  The Hawking process is a symmetric process for globally charged particles - for example particles and anti-particles radiate equally.  To radiate a charge one must have a structure that can induce a chemical potential difference between particles of different charge.  Therefore one needs a gauge field that couples to the particles to generate the necessary chemical potentials (see e.g.~\cite{Hook:2014mla} for a discussion of this perspective).  Hence black hole physics implies that not only are there no continuous global symmetries, but that symmetries in (bulk) quantum gravity must be local gauge symmetries. More precisely in holographic terms, if the dual boundary quantum field theory possesses a splittable global symmetry, the bulk quantum gravity theory must possess a long range gauge symmetry rather than a global symmetry~\cite{Harlow_Main, Harlow_PRL}.

As we have seen above, Ho\v{r}ava-Lifshitz gravity possesses black holes which radiate and have corresponding first laws. As well, it is obvious that the theory is meant to be a consistent theory of quantum gravity. However, given the existence of the reparameterization symmetry outlined in Section~\ref{sec:symmetries}, it is clear that there are symmetries which are not local gauge symmetries but rather mixed. In other words, for Ho\v{r}ava-Lifshitz gravity, the absence of global symmetries does not automatically imply that the symmetries are local gauge. Indeed as we have seen, the existence of a mixed symmetry is actually \textit{required} to resolve outstanding problems with the thermodynamics. This may seem to generate the same contradiction that arose in the heuristic argument above. However, since the equation of motion governing the lapse $N$ is elliptic, as seen in~\eqref{elliptic mode general}, this seems to give a mechanism of generating the appropriate chemical potentials for the black holes to dissipate the elliptic charge and avoid the contradiction found in the heuristic argument. Furthermore, the promise for a full holographic description of Ho\v{r}ava-Lifshitz gravity with boundary Lifshitz field theories~\cite{Horava_Lifshitz, Horava_Membranes, Iranian, CFTno, wedges} begs the question of how it may avoid the existing holographic proofs that bulk symmetries must be long range gauge as found in~\cite{Harlow_PRL,Harlow_Main}. 

\subsubsection{Splittability}

In AdS/CFT, it was proven explicitly in \cite{Harlow_Main, Harlow_PRL} that global symmetries were disallowed in the bulk for quantum gravitational theories. Specifically, it was demonstrated that if a boundary CFT possesses a global symmetry, the symmetry which is dual in the bulk is a long-range gauge symmetry. The proof relied heavily on the assumed \textit{splittability} of global symmetries in boundary CFTs. A splittable global symmetry is roughly defined in the following fashion: Let $U(g,\partial\Sigma)$ be the unitary operator which corresponds to an element $g\in G$ of the symmetry group acting on operators with support on a boundary time slice by conjugation. The global symmetry is said to be \textit{splittable} if the operators $U(g,\partial\Sigma)$ are able to be written as the product of unitary operators with support only on a subset $R_i\subset \partial\Sigma$ where the subsets $R_i$ partition the boundary timeslice $\partial\Sigma$:

\begin{equation}\label{splittability defn}
U(g,\partial\Sigma)=\bigotimes_i U(g,R_i).
\end{equation}

The proof given in \cite{Harlow_Main, Harlow_PRL} began by assuming that a global symmetry on the boundary implied a global symmetry in the bulk. If one then uses entanglement wedge reconstruction, one can construct global symmetry operators in the bulk $U(g,W[R_i])$ with support on the entanglement wedges $W[R_i]$ corresponding to the boundary timeslice regions $R_i$. However, due to the finite mode speeds in AdS/CFT, one can construct a bulk field $\phi$ with support in the ``center'' of the bulk that has empty intersection with all of the $W[R_i]$, implying that $U(g,W[R_i])$ will commute with $\phi$ and that the action of the global symmetry group $G$ on $\phi$ would be trivial for all $g\in G$, contradicting the assumption. In principle, this result should generalize to holographic theories with a notion of entanglement wedge reconstruction.

Since the mixed symmetry is not gauged along a slice $\Sigma$, and since there is a notion of entanglement wedges in Ho\v{r}ava Lifshitz gravity \cite{wedges}, how does Ho\v{r}ava gravity avoid the proof given in \cite{Harlow_Main, Harlow_PRL} if Ho\v{r}ava gravity is holographic? The answer is that the mixed symmetry is not a splittable symmetry: the assumption upon which the proof in AdS/CFT was built. This does not allow one to write the global symmetry operator $U(g,\partial\Sigma)$ of the boundary theory as a product of operators on subregions in the first place, so one can never construct a bulk field $\phi$ which will commute with bulk symmetry operators with support on entanglement wedges corresponding to strict subsets $R_i\subset \partial\Sigma$.

\subsection{Elliptic mode cosmic censorship as a motivator for universal horizons}\label{Censorship Section}

One of the characteristic properties of a universal horizon is its asymptotic nature. Specifically, leaves in the foliation in the past with respect to the \aether time asymptotically approach the universal horizon as $r\to r_{uh}$, or as one goes to the infinite past in metric time along a given slice. This is illustrated by Figure \ref{leaves}, and by \eqref{const T difeq} from which Figure \ref{leaves} is obtained.

There are two ways of interpreting the fact that slices must asymptotically approach the universal horizon. First, one can see that a given slice cannot cross the universal horizon in order to retain causality. Since the universal horizon is a leaf in the preferred foliation, another leaf crossing the universal horizon would generate a contradiction since the foliation, $\phi:\mathcal{M}\to\R\times S$, must be a bijection. The requirement that $\phi$ be a bijection prevents one point $p\in\mathcal{M}$ from being assigned two different \aether times $T,T'$ and ensures that $T$ is a well defined time coordinate that respects causality. A leaf in the foliation crossing the universal horizon (another leaf) would generate this exact contradiction of the injectivity of $\phi^{-1}$.

Another lens through which to look at the asymptotic nature of the universal horizon is that of cosmic censorship. In order for there to be a well defined first law and thermodynamics of universal horizons, leaves in the foliation outside of the universal horizon must in some sense ``reach'' the universal horizon in order to communicate degrees of freedom at the boundary defined by the horizon out to infinity. In the relativistic case, one constructs a surface that intersects the causal horizon and extends out to infinity in order to construct the horizon's first law \cite{Hawking_4_Laws}. As discussed previously, this is disallowed in Ho\v{r}ava-Lifshitz gravity due to the fact that the universal horizon is a leaf in the foliation. 

However, one can come to the same conclusion without knowing that the universal horizon is a leaf, and simply considering the elliptic equation for the lapse given in \eqref{elliptic mode general}, as long as one takes the \textit{additional} step of recognizing that the elliptic mode can carry data. If one were to allow leaves of the foliation to cross the universal horizon, as in the Hawking construction of the first law, by virtue of \eqref{elliptic mode general} being elliptic, specification of initial data at any point outside of the universal horizon would allow for one to solve for the lapse inside of the universal horizon and vice versa, contradicting the fact that $u\cdot\chi=0$ defines a causal horizon. The fact then that leaves must not cross the universal horizon to ensure cosmic censorship while also ``reaching'' the universal horizon (in the weakest sense of the word) in order to construct a first law requires that leaves in the bulk asymptote to the universal horizon as $r\to r_{uh}$ and $v\to-\infty$ along a slice.  Since this is independent of any bulk propagation, instead of requiring Lifshitz behavior for renormalization purposes, an alternative logical path assuming cosmic censorship holds is that the existence of a dynamical foliation requires an elliptic mode, therefore cosmic censorship requires the leaves to asymptote to any putative causal horizon, therefore the quantum gravity theory compatible with this structure must have ultraviolet Lifshitz behavior for bulk modes to make that putative horizon the universal causal horizon for all modes.

\subsection{Future directions}

So far we have established that the elliptic charge, which has generally been ignored in calculations of universal horizon thermodynamics, actually plays an important role.  Requiring it be constant resolves the failure of thermodynamics in asymptotically AdS, but only insofar as it explains why the offending term should be set to zero.  Going forward, the next step is to examine changes in the elliptic charge and how they appear in the first law, just like other charges do.  Doing so requires both an understanding of the variation of the elliptic data, i.e. the boundary conditions for the elliptic mode, but also an understanding of the mechanism by which elliptic data at the horizon could be emitted in a Hawking process.  This would be a novel extension of black hole thermodynamics for information that does not ``propagate out'' through the bulk, but instead directly couples information on the boundary with horizon information.  We leave this for future work.

\acknowledgments
We are grateful to Stefano Liberati for reading a draft of this work and providing useful feedback.  LM and DM thank the University of New Hampshire and the US Department of Energy under DOE grant DE-SC0020220 for support while developing these ideas.


\bibliographystyle{JHEP}
\bibliography{bib}




\end{document}